\documentclass[preprint]{elsarticle}
\usepackage{amsmath}
\usepackage{graphicx}
\usepackage{subfigure}
\usepackage{calc}
\usepackage{color}
\usepackage{xcolor}
\usepackage{amssymb}
\usepackage[figuresright]{rotating}
\begin{document}
\setlength{\parindent}{0.5cm}
\setlength{\parskip}{0.0cm}
\biboptions{sort&compress}
\bibliographystyle{elsarticle-num}
\begin{frontmatter}

\title{The Blume-Capel Model on Hierarchical Lattices: exact local properties}
\author[dfufpe]{M\'ario  J. G. Rocha-Neto}
\ead{mario.neto@garanhuns.ifpe.edu.br}
\author[caufpe]{G. Camelo-Neto}
\ead{gustavo.camelont@ufpe.br}
\author[dfufpb]{E. Nogueira Jr.}
\ead{enogue@fisica.ufpb.br}
\author[dfufpe]{S. Coutinho\corref{cor1}}
\ead{sergio@ufpe.br}
\cortext[cor1]{Corresponding author}
\address[dfufpe]{Laborat\'orio de F\'{\i}sica Te\'orica e Computacional,
Departamento de F\'{\i}sica,
Universidade Federal de Pernambuco,
50670-901, Recife, Brazil}
\address[caufpe]{N\'ucleo Interdisciplinar de Ci\^encias Exatas e da Natureza, Universidade Federal de Pernambuco, 55.014-900, Caruaru, Brazil.}
\address[dfufpb]{Departamento de F\'isica, Universidade Federal da Para\'iba, 58051-970, Jo\~ao Pessoa, Brazil.}
\date{\today}
\begin{abstract}
The local properties of the spin one ferromagnetic Blume-Capel model defined on hierarchical lattices with dimension two and three are obtained by a numerical recursion procedure and studied as functions of the temperature and the reduced crystal-field parameter. The magnetization and the density of sites in the configuration $S = 0$ state are carefully investigated at low temperature in the region of the phase diagram that presents the phenomenon of phase reentrance. Both order parameters undergo transitions from the ferromagnetic to the \emph{ordered} paramagnetic phase with abrupt discontinuities that decrease along the phase boundary at low temperatures. The distribution of magnetization in a typical \emph{profile} was  determined on the transition line presenting a broad multifractal spectrum that narrows towards the fractal limit (single point) as the discontinuities of the order parameters grow towards a maximum. The amplitude of the order-parameter discontinuities and the narrowing of the multifractal spectra were used to delimit the low temperature interval for the possible locus of the tricritical point.
\end{abstract}
\end{frontmatter}
\section{Introduction}\label{section0}
Since the second half of the past century, spin models have become a powerful tools for studying phase transitions in various physical systems beyond magnetism. The spin-one Ising model formulated independently by Blume\cite{blume66} and Capel\cite{capel66}, hereafter Blume-Capel model (BC), is one of the most investigated since then. It is the precursor spin model in the study of first-order phase transitions to include an anisotropic crystal field interaction in addition to the bilinear exchange interaction between the neighboring spins. Another seminal spin-one Ising model is the Blume-Emery-Griffiths model\cite{blume71} (BEG), which generalizes the BC model and was  successfully used to simulate and describe the thermodynamic behavior of the superfluid-normal phase separation in the liquid $^3$He -$^4$He mixtures along the $\lambda$-line and in the neighborhood of the critical point. Both models and their variations were studied by several theoretical methods, experimental techniques and computational simulations in Statistical and Condensed Matter Physics. 

The BC model was first studied by molecular field approximation (MFA)\cite{blume66, capel66, plascak93}, and later explored by other methods and techniques beyond MFA such as the effective field approximations (EFA)\cite{siqueira86,bonfim85,hoston91,polat03,costabile12,costabile14} and the cluster variation methods (CVM)\cite{buzano95,balcerzak04,erhan13}. It was also investigated by high- and low-temperatures series expansions\cite{oitmaa70, oitmaa72,saul74}, Monte Carlo (MC) simulations\cite{jain80,landau86,kimel87,puha01,silva06,kwak15} and by position-space renormalization group\cite{berker76,branco97,bouziani13}. Some variants of the BC model with other spin variables and the presence of disorder in the fields and/or interactions have been objecting of recent studies \cite{amadeu00,ekiz01,ozkan06,snowman11,yuksel12,madani15,erhan16,santos17}. The existence of a tricritical point (TCP) has been detected in the vicinity of the boundary separating regions with non-zero magnetization from that with zero magnetization through the MFA/EFA solutions and computational simulation. The position of the tricritical point depends greatly on the method used to investigate the phase transition but is still an object of interest. For instance, the TCP locus of the \emph{emblematic} square-lattice BC model has been searched by accurate Monte Carlo computational simulation and transfer matrix methods. A recent review by Zierenberg and coauthors \cite{zierenberg17} gives several close values of its location obtained by such methodologies.

In this work, we present the exact solution of the BC model defined on diamond hierarchical lattices (DHL). Hierarchical lattices in spite of the lack of translational symmetry possess full-scale invariance symmetry and have been widely used to exactly study spin models\cite{kaufman81, griffiths82,melrose83,kaufman84}. In particular, the solution of the Ising model on diamond hierarchical lattices has been shown to be equivalent to the Migdal--Kadanoff renormalization group approximation\cite{migdal75,kadanoff76} for corresponding Bravais lattices \cite{bleher79,berker79}. Here we combine an exact real-space renormalization group scheme with an exact recursive procedure to  get the local properties (site-by-site) of the model as functions of the temperature and fields.

The main aim of this work is to investigate in detail the behavior of the densities of sites, in which the spin variable is in the state $S = 0$ or $S = \pm 1$ at low temperatures. Particularly in the vicinity of the boundary separating regions with non-zero magnetization from that with zero magnetization, the latter with the majority of sites in the $S=0$ state. The presence  of a tricritical point has been detected at this border through the MFA/EFT solutions and computational simulation, however, the characteristics of the spins state configurations in this region have not been completely explored.

The combined procedure to get the local properties of the model was originally proposed to study the local magnetization of the ferromagnetic Ising model\cite{morgado90, morgado91,coutinho92} and later applied to local order parameter of the Ising spin-glass model\cite{nogueira97}, of the random-field Ising model\cite{rosas04} as well as of the q-state Potts model\cite{silva96}. In these works, the local order parameter shows a multifractal structure in the vicinity of the continuous phase transition, resulting from the combination of the fractal nature of the diamond hierarchical lattice with the critical behavior of the order parameter. Moreover, it was observed that the spectrum of the fractal measure defined by the order parameter is directly related to its critical exponent\cite{morgado91,coutinho92, silva96}. 

In the following section the phase diagrams of the BC model on diamond hierarchical lattices of dimension $d = 2$ and $3$ are obtained and discussed. In section \ref{section3}, the general recursive procedure to obtain the local properties of the BC model under field  on hierarchical lattices with arbitrary dimension is presented. The order parameters for the ferromagnetic phase (magnetization) and for the \emph{ordered} paramagnetic phase (density of $S=0$ spins) are presented and discussed as functions of the temperature and the crystal-field strength. Section \ref{section4} is devoted to investigate and discuss the multifractal properties of the local magnetization  in the vicinity of the phase transitions observed in the model. Finally, in section \ref{section5} a summary of the main conclusions is presented.

\section{The Hamiltonian model and phase diagram.}\label{section2}
The general reduced Hamiltonian $\mathcal{H}_{\textsf{BC}}$ for the Blume-Capel model can be written as
\begin{equation}\label{eq1}
-\beta \mathcal{H}_{\textsf{BC}}=\sum_{\langle ij\rangle} J_{ij} S_iS_j- \sum_{i}\Delta_i S_i^2 +  \sum_{i}h_i S_i,
\end{equation}
where $\beta=1/k_B T$, $k_B$ is the Boltzmann constant and $T$ the absolute temperature. $J_{ij}$ are the reduced exchange coupling constants interactions associated with the nearest-neighbor spins pairs $\langle ij \rangle$, $\Delta_i$ represents the reduced  local crystal field and $h_i$ is the reduced  local magnetic field both acting on the spin site $S_i$ with variables $S_i=0,\pm 1$.

\begin{figure}[ht]
\begin{center}
\includegraphics[width=8cm]{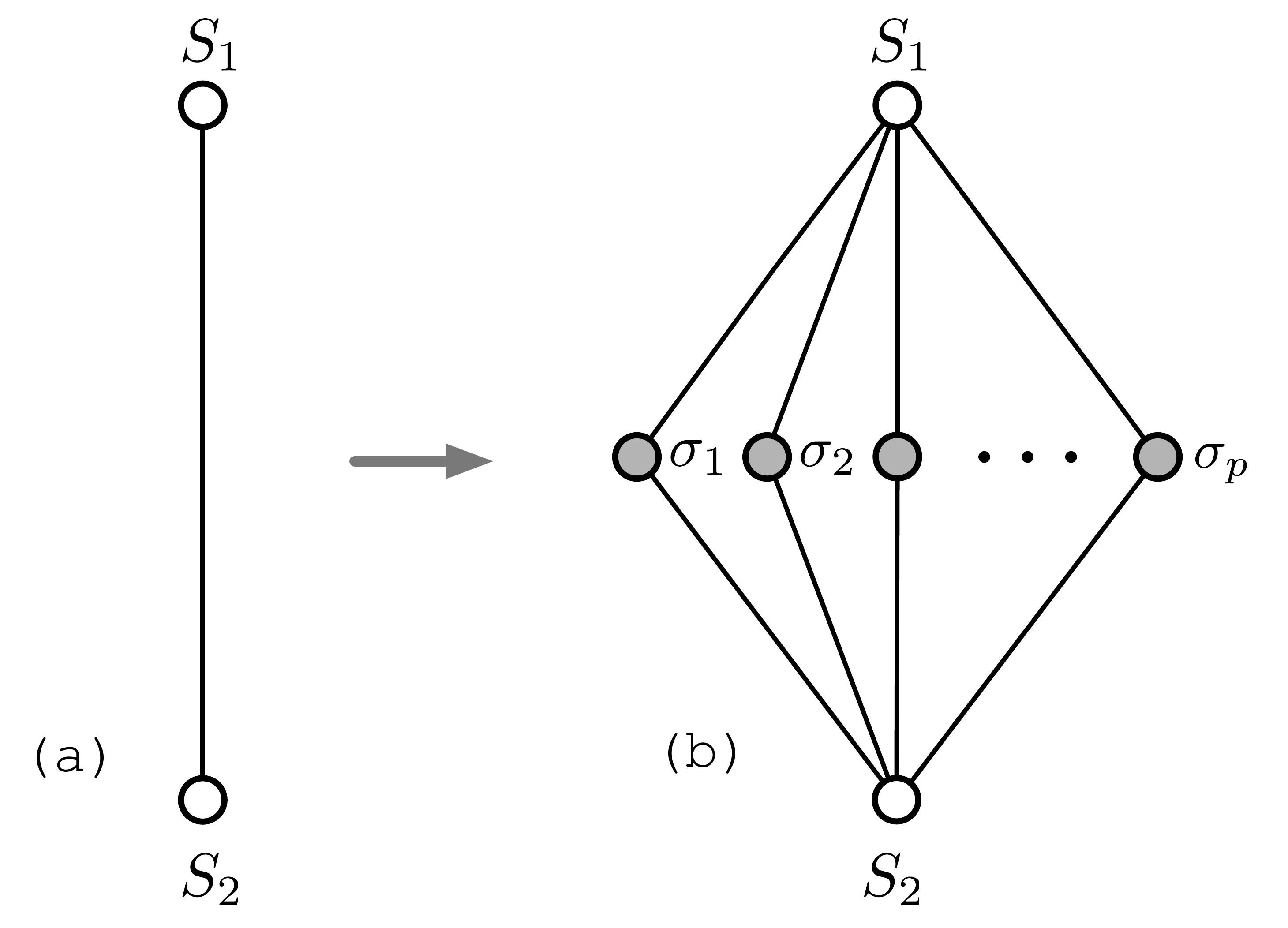}
\caption{Diamond hierarchical lattice with scaling factor $b=2$ and $p$ connections: (a) zeroth generation (b) first generation or basic unit. The gray arrow indicates the inflation process. Open circles denote the root sites $S_1$ and $S_2$ and the solid gray circles label the internal ones $\sigma_j$ ($j=1,\dots p$). The scaling factor $b=a+1$, where $a$ is the number of internal sites within each connection.}
\label{fig1}
\end{center}
\end{figure}

In the present work, the BC model is defined on the diamond family of hierarchical lattices (DHL) of general fractal dimension. Hierarchical lattices are constructed by an \emph{inflation} process \cite{kaufman81,griffiths82}, the zeroth generation being composed by a single bond joining the root sites, which is successively replaced by the basic unit that characterizes the lattice topology. After $n$ steps, this inflation procedure leads to an $n$-generation two-rooted lattice. The DHL basic unit with scaling factor $b=2$ and $p$ connections, which is illustrated in Figure \ref{fig1}, generates lattices with fractal dimension $d=1+\ln p/\ln 2$ in the limit of infinite generations. For a finite $n$-generation lattice there are $N_s(n,p)= 2+p[(2p)^n-1]/(2p -1)$ sites and $N_b(n,p)=(2p)^n$ bonds, for which there are $p(2p)^{\ell-1}$ sites introduced at the generation $\ell$ ($\ell \leq n$), whose \emph{coordination number} is $2p^{n-\ell}$. 

The BC model defined on the DHL of scaling factor $b=2$, with $n$ generations can be exactly transformed under real-space renormalization group procedure in an equivalent model defined on DHL with $(n-1)$ generations provide the $p$ internal spins variables within each basic unit are decimated inverting the arrow in the Figure \ref{fig1}. Afterwards, the partition functions for both models are matched, less than a generation dependent multiplicative factor. Such \emph{decimation} process corresponds in a certain sense to a \emph{deflation} of the $n$-generation hierarchical lattice  into an $(n-1)$ one with a new set of coupling constants. Under such renormalization procedure, the new set of coupling constants, here indicated by the prime superscript, is obtained by a formal set of renormalization equations.  For the Blume-Capel Hamiltonian given by equation (\ref{eq1}) on a DHL, these equations can be written as
\begin{align}
J' &=\frac{1}{4}\sum_{i=1}^p \ln \left[\frac{\Phi_i[1,1]\Phi_i[-1,-1]}{\Phi_i[1,-1]\Phi_i[-1,1]}\right], \label{eq7}\\
\Delta'_1& =\Delta_1-\frac{1}{2} \sum_{i=1}^p \ln \left[\frac{\Phi_i[1,0]\Phi_i[-1,0]}{\Phi_i[0,0]\Phi_i[0,0]}\right],\\
\Delta'_2& =\Delta_2-\frac{1}{2} \sum_{i=1}^p \ln \left[\frac{\Phi_i[0,1]\Phi_i[0,-1]}{\Phi_i[0,0]\Phi_i[0,0]}\right],\\
h_1'&= h_1+ \frac{1}{4} \sum_{i=1}^p \ln \left[\frac{\Phi_i[1,1]\Phi_i[1,-1]}{\Phi_i[-1,-1]\Phi_i[-1,1]}\right],\\
h_2'&= h_2+ \frac{1}{4} \sum_{i=1}^p \ln \left[\frac{\Phi_i[1,1]\Phi_i[-1,1]}{\Phi_i[-1,-1]\Phi_i[1,-1]}\right],\label{eq10}
\end{align}
with
\begin{align}
\label{eq11}
\Phi_i[S_1,S_2]&=\textsf{Tr}_{\{\sigma_i\}} \exp{[J_{1i}\sigma_i S_1 + J_{2i} \sigma_i  S_2-\Delta_i \sigma_i^2+h_i \sigma_i]=}\nonumber \\
& = 1+ 2e^{-\Delta_i}\cosh[J_{1i}S_1+J_{2i} S_2 +h_i]
\end{align}
and $\{\sigma_i\}$ indicating the set of configurations of the internal spin variables which were decimated. 

\begin{figure}[ht]
\begin{center}
\includegraphics[width=10cm]{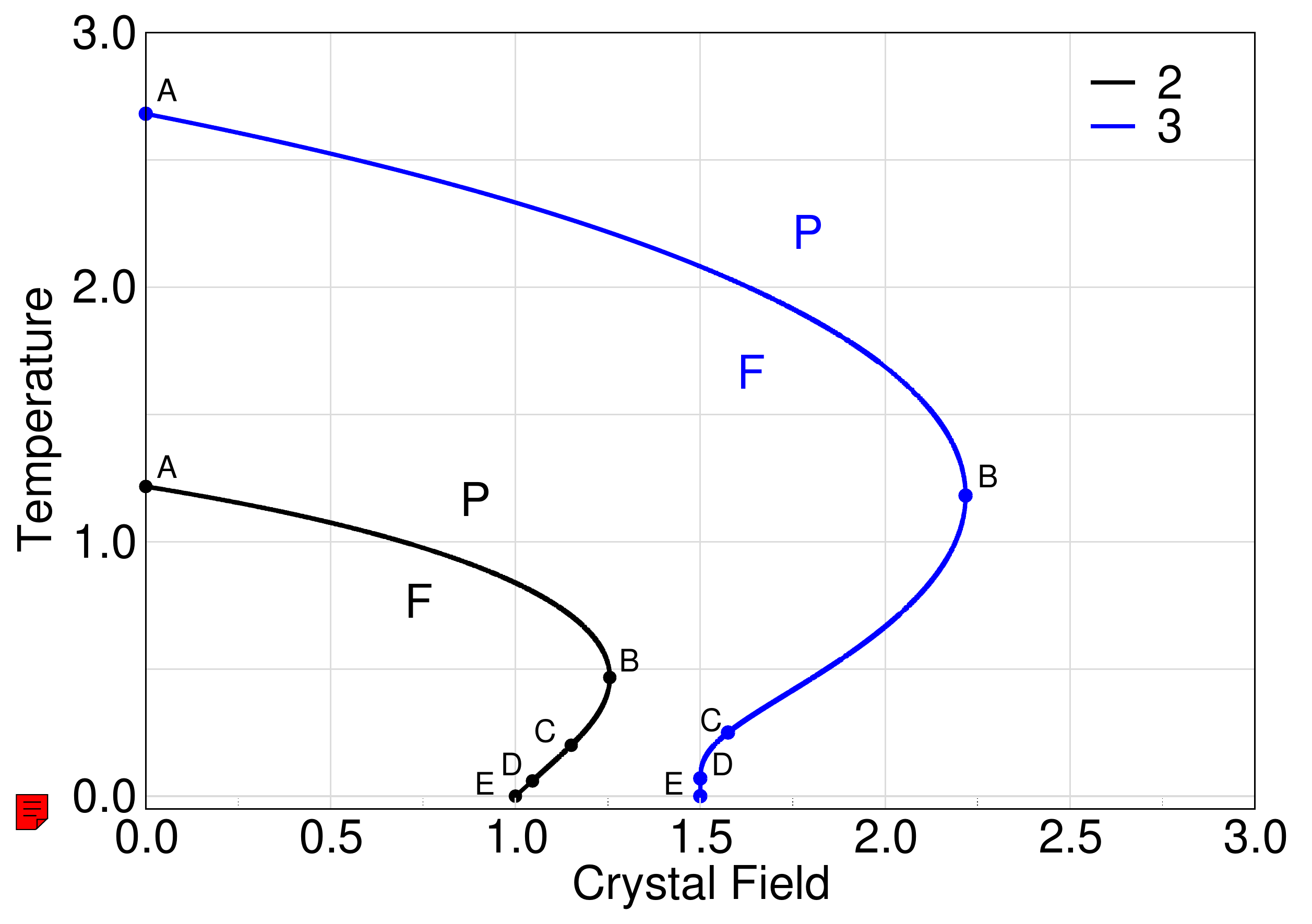}
\caption{Temperature $(1/J$) \emph{versus} crystal field $(\Delta/J)$ phase diagram for the Blume-Capel model on a DHL of  dimension $d=2$ and $3$. F indicates the ferromagnetic region, and P labels the paramagnetic one for the respective dimensions: black for $d=2$  and blue for $d=3$.  Letters A, B, C, D and E indicate the relevant points shown in Table 1 below. The unstable fixed points ($T_{I}, \alpha_{I}=-\infty$) for each case are not displayed in the diagram.}
\label{fig2}
\end{center}
\end{figure}

In the zero-field regime ($h_i=0$) and uniform exchange interactions ($J_{ij}=J$), the above set of renormalization equations reduces to 
  \begin{align}
\label{eq12}
   J'&=\ln \left[\frac{1+ 2 e^{-\Delta} \cosh(2 J)}{1+ 2 e^{-\Delta}}\right]^{p/2} \mbox{,}    \\
   \Delta' &= \Delta - \ln \left [\frac{1+ 2 e^{-\Delta} \cosh( J)}{1+ 2 e^{-\Delta}}\right]^p \mbox{.}  \label{eq13}
\end{align}
The renormalization flow within the $(J,\Delta)$-parameter space governed by equations (\ref{eq12}) and  (\ref{eq13}) is characterized by two stable fixed-points: the ferromagnetic one at ($J \to \infty, \Delta \to -\infty$) with $\Delta/J \to -p/(p-1)$ and a paramagnetic line of fixed-points at ($J=0,  -\infty \leq\Delta \leq \infty$). An unstable fixed point is also found in ($J^* , \Delta =-\infty$).
The $\Delta \to -\infty$ manifold in the parameter space corresponds to the spin 1/2 Ising model limit with critical temperature given by $T_{I} = 1/J^*$. The \emph{locus} of the line separating these two basins of attraction ends in this unstable fixed point and can be numerically drawn being better represented in the appropriated temperature $(T=1/J$) \emph{versus} crystal field $(\alpha=\Delta/J)$ parameter space, which plays to role of the phase diagram of the BC model, as usual. Figure \ref{fig2} displays the particular plots of these diagrams for the BC model defined on DHL's of integer dimension $d=2$ ($p=2$) and $d=3$ ($p=4$). 

The renormalization flow  in the $T \times \alpha$ parameter space is also characterized by two basins of attraction: the ferromagnetic sink at low temperatures governed by the stable fixed point at $(0, \alpha^*_{\textsf{F}})$ and the high temperature paramagnetic one within which the flow is attracted towards the stable fixed point at $(\infty, -\infty)$ not shown in Figure \ref{fig2}. 
At zero temperature the renormalization equations given by (\ref{eq12}) and (\ref{eq13})  reduce to
\begin{equation}
\label{eq15}
 \alpha'=\begin{cases}
\infty, & \alpha \geq 2, \\[2ex]
   \displaystyle  \frac{2\,\alpha}{p(2-\alpha)}, & 1 <\alpha <2, \\[3ex]
   \displaystyle \frac{2[(1+p)\alpha - p]}{p(2-\alpha)},& 0 <\alpha <1, \\[3ex]
       \displaystyle   \frac{\alpha-p}{p}, & \alpha <0.
\end{cases}
\end{equation}

At the $T=0$ subspace there exist an unstable fixed point at $(0, \alpha^*)$ with $\alpha^* = 2(1-2^{1-d}) $ separating the ferromagnetic phase (F) of configuration $\{S\}=1$ (or $\{S\}=-1$), which is governed by the stable fixed point at $\alpha_{\textsf{F}}^* =- 2^d/(2^d-2)$, and the \emph{ordered} paramagnetic phase (OP) of configuration $\{S=0\}$ whose stable fixed point is located  at ($0,\infty$). Note that for a fixed value of $\alpha$ just above $\alpha^* $ and non-zero temperatures, there is an interval of ($\alpha^*,\alpha_{\max})$  where two possible transitions can occur: para$\to$ferromagnetic followed by ferro$\to$paramagnetic transitions as the temperature is lowered from high temperatures. This interval can be split into two subintervals, $(\alpha^*,\alpha_{\textrm d})$ and $(\alpha_{\textrm d} , \alpha_{\textrm c})$. Within the first one, it was observed that the ferromagnetic--paramagnetic transition exhibit a full discontinuity in the order parameters and displays a \emph{fractal profile} while within the later the discontinuity in the ferromagnetic--paramagnetic transition vanishes monotonically as $\alpha$ grows and the profile patterns become \emph{multifractal} as will be discussed in sections  \ref{section3} and \ref{section4}. Table 1 provides the locus of these particular points for the model defined on lattices with $d=2$ and 3, that will be of interest later. 

\begin{table}[ht]
\renewcommand{\arraystretch}{1.5}
\setlength{\tabcolsep}{7pt}
  \centering 
  \label{tab}
\begin{tabular}{|c|ll|cc|cc|}
\hline
\multicolumn{1}{c}{}&\multicolumn{2}{c}{}  & \multicolumn{2}{c}{d=2 }& \multicolumn{2}{c}{ d=3 } \\ \hline
I& $T_\textrm I$& $-\infty $ &  $1.641$ & $ -\infty$& $3.830$ & $-\infty$ \\ \hline
A& $T_0$ & $ \alpha_0$  & $1.217 $ & $ 0.000$   & $ 2.682$ & $0.000$ \\ \hline
 B& $T_{\alpha_{\max}}$ & $\alpha_{\max}$ & $ 0.466$ & $1.255$ & $1.181$ & $2.218$ \\ \hline
  C&  $T_\textrm c$ & $\alpha_\textrm c$&$0.200$ & $1.151$ & $0.250$ & $1.575$ \\ \hline
 D&  $T_\textrm d $ & $\alpha_\textrm d$&$0.060$ & $1.046$ & $0.070$ & $1.500$ \\ \hline
E &   $T=0$ & $\alpha^*$ &0 & 1.000& 0 & 1.500\\
\hline
\end{tabular}
  \caption{Coordinates of the relevant points in the phase diagram for the model in a DHL with dimensions 2 and 3: I=($T_\textrm I,\,-\infty$) spin 1/2 Ising model limit; A=($T_0,\; \alpha_0=0$) zero crystal-field  BC model; B=($T_{\alpha_{\max}},\; \alpha_{\max}$) threshold point for absence of ferromagnetic phase  when $\alpha > \alpha_{\max}$; C=($T_{\textrm c},\; \alpha_{\textrm c}$) threshold point for a continuous phase transition for $T\geq T_{\textrm c}$ and D=($T_{\textrm d},\; \alpha_{\textrm d}$) threshold point for existence of phase transition with full discontinuity in the order parameters for $ T<T_{\textrm d}$ as explained in text below and D=($T=0,\; \alpha^*$) critical field for the zero temperature ferromagnetic-\emph{condensed} paramagnetic transition in the ground state.}
\end{table}

In the following sections, the nature of the F and P phases in the vicinity of such transition points will be explored by means of the temperature and $\alpha$ dependent local properties obtained by an exact recursion procedure, which is presented in the following section. 
\section{Exact local properties}
\label{section3}
In this section and hereafter, the study is focused on the ferromagnetic BC model defined on diamond hierarchical lattices with scaling factor $b=2$ and $p$ connections. The generalization for other values of $b$ is straightforward but leading to very much involved algebraic equations without relevant benefits to the study of the pure ferromagnetic case. Here we generalize to the BC model ($S=1$) the method proposed in the reference \cite{morgado90} to investigate the local properties of the $S=1/2$ Ising model on DHL's.

\begin{figure}[ht]
\begin{center}
\includegraphics[width=6cm]{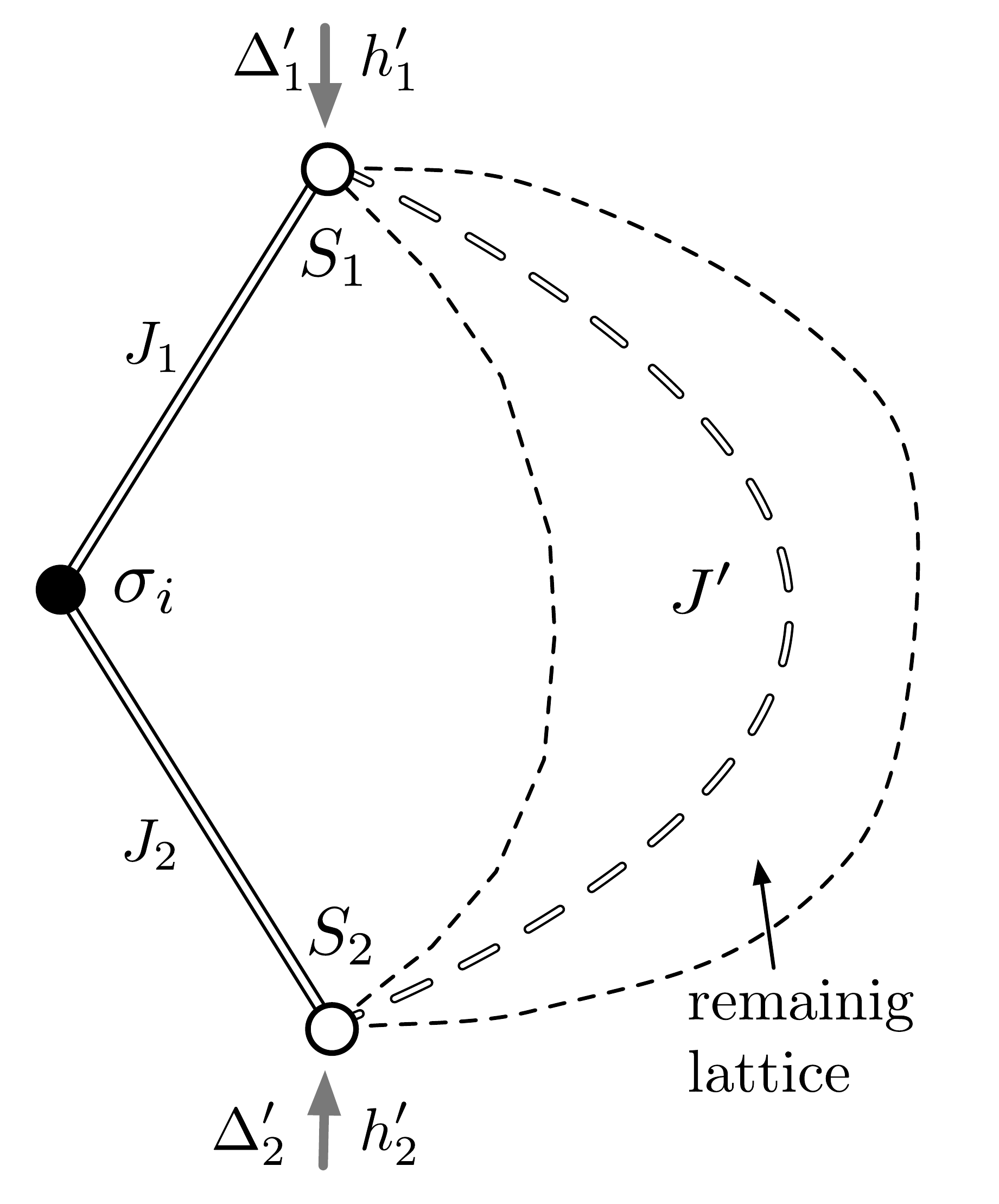}
\caption{Schematic chart for the effective model. Double line represents the interactions ($J_i$) between the spin of the inner site (black bullet) and the ones at the root sites (white circles) of a single connection of an arbitrary basic unit. Double dashed line represents  the effective interactions ($J'$) due to the whole DHL, except those of the considered connection ($i=1,2$). Solid gray arrows represent the effective external magnetic field and single-ion anisotropy field.}
\label{fig3}
\end{center}
\end{figure}

The temperature reduced Hamiltonian for the BC model given by equation (\ref{eq1}) and defined on an $n$-generation DHL can be decomposed into two parts: $\mathcal{H}_\textsf{int}$ referring to a single connection of a particular basic unit introduced in the last generation and $\mathcal{H}_\textsf{ext}$ which comprises all other remaining interactions, i.e.,
\begin{equation}\label{eq2}
\mathcal{H}=\mathcal{H}_\textsf{int}+\mathcal{H}_\textsf{ext} \mbox{,}
\end{equation}
where
\begin{align}\label{eq3}
-\beta \mathcal{H}_\textsf{int}=&J_{1i}\,\sigma_i S_1+J_{2i}\,\sigma_i  S_2-\Delta_i\sigma_i^2+h_i\sigma_i \,\mbox{,}
\end{align}
and
\begin{equation}\label{eq4}
-\beta \mathcal{H}_\textsf{ext}= J'S_1S_2-\Delta'_{1}S_1^2-\Delta'_{2}S_2^2+h'_{1}S_1+h'_{2}S_2 \mbox{.}
\end{equation}
In  equation (\ref{eq3}) $J_{ji}$ ($j=1,\,2$) are the ferromagnetic exchange coupling constants associated with the pair interactions of the \emph{internal} spin variable $\sigma_i$ with the \emph{external} spin variable $S_j$ ($j=1,2$), respectively, and $h_i$ is the external field acting of all spins. On the other hand equation (\ref{eq4}) describes the effective Hamiltonian of the remaining lattice variables where the unknown \emph{effective} couplings and fields  $J'$, $\Delta'_\ell$ and $h_\ell'$ ($\ell=1,\,2$)  act on the external spin variables $S_j$, ($j=1,2$) as illustrated in Figure \ref{fig3}. Such representation was inspired in the calculation of the effective interaction of \emph{decorated} spin models\cite{fisher59} where the effective fields and interactions are unknown parameters to be determined.\\

The partition function for this effective Hamiltonian can be conveniently written as
\begin{equation}
\label{eq5}
\mathcal Z =\textsf{Tr}_{\{S_1,S_2\}} Z[S_1,S_2] \Phi[S_1,S_2]\, ,
\end{equation}
where
\begin{align}
\label{eq6}
   \Phi[S_1,S_2]  &= \textsf{Tr}_{\{\sigma_i\}} \exp{[ -\beta \mathcal{H}_\textsf{int}]}, \\[1ex]
    Z[S_1,S_2] &=  \exp{[-\beta \mathcal{H}_\textsf{ext}]}.
\end{align}
Now, the partition function can be expanded according to the nine possible  configurations of the $\{S_1, S_2\}$
\begin{equation}
\label{eq9}
\mathcal Z = \sum_{j=1}^9 \Phi_j \, X_j\,,
\end{equation}
where the ${\Phi_j}$'s are known quantities $\Phi[S_1,S_2]$ while the $X_j$'s indicate the unknown quantities $Z[S_1,S_2]$ to be determined. 

The local properties of interest are the site-magnetization $m_i=\langle \sigma_i \rangle$, the quadrupolar parameter $q_i=\langle \sigma^2_i \rangle$ and the correlations $\langle \sigma_i \sigma_j\rangle $, $\langle \sigma^2_i \sigma_j \rangle $, and $\langle \sigma^2_i \sigma_j^2\rangle $. 

The procedure to obtain these quantities can be summarized in the following steps:
\begin{description}
\item[(a)] Calculate $\langle \sigma_i \rangle$, $\langle \sigma^2_i \rangle$, $\langle \sigma_i S_{1}\rangle $,  $\langle \sigma_i S_{2}\rangle $, $\langle \sigma^2_i S_{1}\rangle $, $\langle \sigma^2_i S_{2}\rangle $, $\langle \sigma_i S^2_{1}\rangle $,  $\langle \sigma_i S^2_{2}\rangle $, $\langle \sigma^2_i S^2_{1}\rangle $ and $\langle \sigma^2_i S^2_{2}\rangle $ associated with the internal site $\sigma_i$, which will be linear functions of the unknown nine quantities  $X_j$'s. 
\item[(b)] Calculate $\langle S_1 \rangle$,   $\langle S_2 \rangle$, $\langle S_1^2 \rangle$, $\langle S_2^2 \rangle$, $\langle S_1 S_{2}\rangle $, $\langle S^2_1 S_{2}\rangle $, $\langle S^2_2 S_{1}\rangle $ and $\langle S^2_1 S^2_{2}\rangle $ associated with the root sites $S_j$, which will be also linear functions of the unknown quantities $X_j$'s. 
\item[(c)] The system of the eight linear equations calculated in (b) can be inverted with help of the partition function  giving the unknown nine quantities $X_j$'s as functions of the \emph{external} local quantities, which can now be substituted into the equations for quantities calculated in (a) yielding after some algebraic manipulations into two independent systems of recursion equations relating local quantities belonging to successive hierarchies of the lattice, given below.
\end{description}
\begin{flalign} \label{eq22}
\langle \sigma  \rangle_{n+1} =W^{(n)}_1[\langle S_1 \rangle_{n}& +\langle S_2 \rangle_{n} ]-[W^{(n)}_1- W^{(n)}_2]\nonumber
\\  & \qquad \qquad \qquad {}\times [\langle S_1^2 S_2 \rangle_{n} +\langle S_1S_2^2 \rangle_{n} ],\\[1ex]
\langle\sigma S_1^2 \rangle_{n+1} =W^{(n)}_1\langle S_1 \rangle_{n}& +W^{(n)}_2 \langle S_1^2 S_2  \rangle_{n} -[W^{(n)}_1- W^{(n)}_2]\langle S_1S_2^2 \rangle_{n}, \\[1ex]
\langle \sigma S_2^2\rangle_{n+1}=W^{(n)}_1\langle S_2 \rangle_{n}& +W^{(n)}_2 \langle S_1S_2^2 \rangle_{n} -[W^{(n)}_1- W^{(n)}_2]\langle S_1^2 S_2 \rangle_{n}, \\[1ex]
\langle \sigma^2S_1 \rangle_{n+1} =W^{(n)}_3\langle S_1 \rangle_{n}& +[W^{(n)}_4-W^{(n)}_3+ W^{(n)}_5]\langle S_1S_2^2 \rangle_{n}  \nonumber \\ 
 &\qquad \qquad \qquad - [W^{(n)}_4- W^{(n)}_5] \langle S_1^2 S_2  \rangle_{n}, \\[1ex]
\langle\sigma^2S_2 \rangle_{n+1} =W^{(n)}_3\langle S_2 \rangle_{n}& +[W^{(n)}_4-W^{(n)}_3+ W^{(n)}_5]\langle S_1^2 S_2 \rangle_{n} \nonumber \\
&\qquad \qquad \qquad  -[W^{(n)}_4- W^{(n)}_5]\langle S_1S_2^2 \rangle_{n},    \label{eq26}
\end{flalign}
for the average of  spin variables of even order, and
\begin{flalign} \label{eq27}
\langle\sigma^2 \rangle_{n+1}&=[W^{(n)}_3-2W^{(n)}_4][\langle S_1^2  \rangle_{n} +\langle S_2^2 \rangle_{n} ]-[W^{(n)}_4- W^{(n)}_5]\langle S_1 S_2 \rangle_{n}\nonumber\\
&\qquad \qquad \qquad+[3W^{(n)}_4-2W^{(n)}_3+ W^{(n)}_5]\langle S_1^2S_2^2 \rangle_{n} +2W^{(n)}_4,\\[1ex]\langle\sigma S_1 \rangle_{n+1} &= W^{(n)}_2\langle S_1 S_2 \rangle_{n} +W^{(n)}_1\langle S_1^2 \rangle_{n} -[W^{(n)}_1- W^{(n)}_2]\langle S_1^2S_2^2 \rangle_{n}, \\[1ex]
\langle\sigma S_2 \rangle_{n+1} &= W^{(n)}_2\langle S_1 S_2 \rangle_{n} +W^{(n)}_1\langle S_2^2 \rangle_{n} -[W^{(n)}_1- W^{(n)}_2]\langle S_1^2S_2^2  \rangle_{n}, \\[1ex]
\langle\sigma^2S_1^2\rangle_{n+1}&=W^{(n)}_3\langle S_1^2 \rangle_{n} -[W^{(n)}_4- W^{(n)}_5]\langle S_1S_2 \rangle_{n} \nonumber \\
& \qquad \qquad \qquad \qquad \qquad +[W^{(n)}_4-W^{(n)}_3+ W^{(n)}_5]\langle S_1^2 S_2^2 \rangle_{n}, \\
\langle\sigma^2S_2^2 \rangle_{n+1} &=W^{(n)}_3\langle S_2^2 \rangle_{n}-[W^{(n)}_4- W^{(n)}_5]\langle S_1S_2 \rangle_{n}  \nonumber \\
& \qquad \qquad \qquad \qquad \qquad \qquad +[W^{(n)}_4-W^{(n)}_3+ W^{(n)}_5]\langle S_1^2 S_2^2 \rangle_{n}. \label{eq31}
\end{flalign}
for the average spin variables of odd order,
where
\begin{align}
\label{}
  W^{(n)}_1  &= \frac{e^{2J^{(n)}}-1}{1+ e^{\Delta^{(n)} + J^{(n)}} + e^{2J^{(n)}}},   \\[1ex]
  W^{(n)}_2  &=\frac{1}{2} \frac{e^{4J^{(n)}}-1}{1+ e^{\Delta^{(n)} + 2J^{(n)}} + e^{4J^{(n)}}}, \\[1ex]
  W^{(n)}_3  &= \frac{e^{2J^{(n)}}+1}{1+ e^{\Delta^{(n)} + J^{(n)}} + e^{2J^{(n)}}},   \\[1ex]
  W^{(n)}_4 & = \frac{1}{2 + e^{\Delta^{(n)}}},\\
   W^{(n)}_5  &=\frac{1}{2} \frac{e^{4J^{(n)}}+1}{1+ e^{\Delta^{(n)} + 2J^{(n)}} + e^{4J^{(n)}}},
\end{align}
the superscript $(n)$ indicating the order of the hierarchy.
\subsection{Magnetization and quadrupolar moment} 
The magnetization \emph{per spin} defined by 
\begin{equation}
\label{eq33}
m=\frac{1}{N_s}\sum_{j=1}^{N_s}\langle \sigma_j \rangle,
\end{equation} 
where here $N_s=N_s(n,p)$ is the total number of spins of a finite DHL with $n$ generations. For fixed values of the temperature $T$ and the crystal-field parameter $\alpha$,  $m$ can be numerically calculated  by summing all local values of $\langle \sigma_i \rangle $ obtained by iterating the system of equations (\ref{eq22})-(\ref{eq26}) for each basic unit cell of the DHL. This iteration process begins with a DHL with $n = 0$ generation (a single bond between the root sites) and the initial values of the quantities corresponding to those that characterize the configuration associated with the stable fixed point of the ordered phase. With help of the recurrence equations (\ref{eq22}-\ref{eq26}), the corresponding values associated with the new sites and bonds incorporated in the DHL with $n = 1$ generations (basic cell) are calculated. Thereafter, such values are conveniently stored and used to obtain the respective values associated with the new sites and bonds incorporated to build the next DHL generation $n = 2$ and so on.

\begin{figure}[htb]
 	\centering
 	\subfigure{%
 	\includegraphics*[width=8cm]{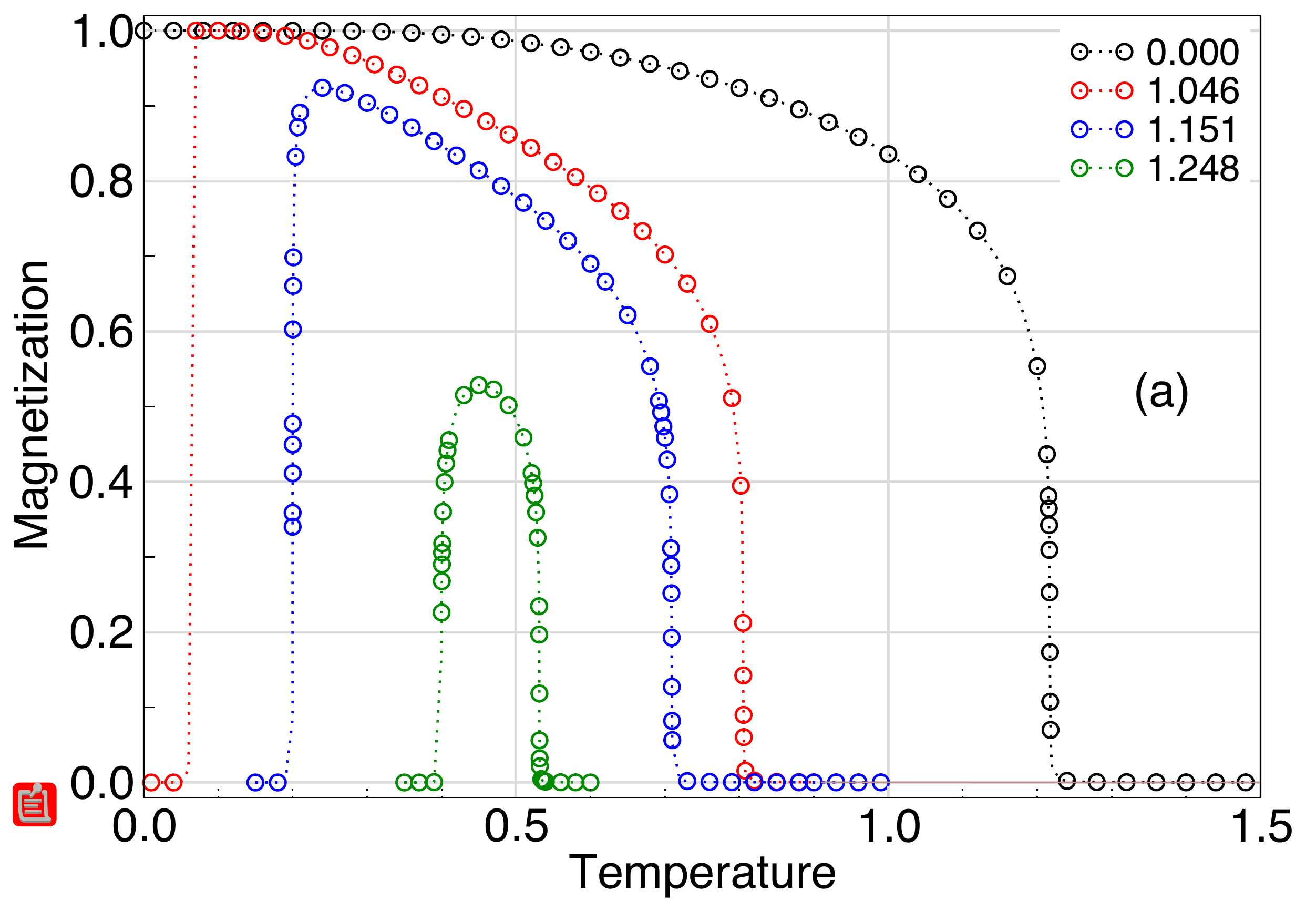}}\quad
 	\subfigure{%
 	\includegraphics*[width=8cm]{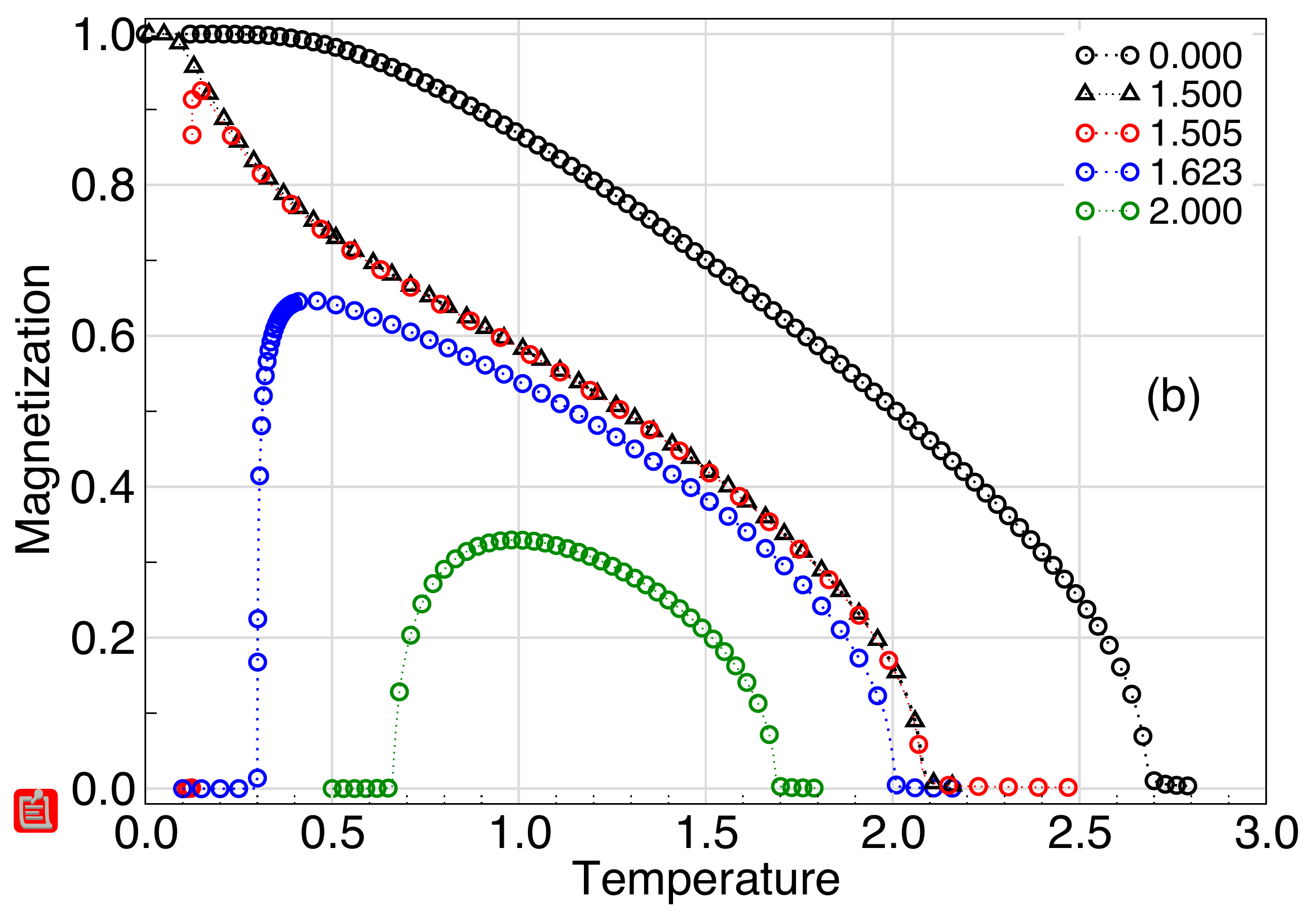}}
 	\caption{Magnetization per spin against temperature for several values of $\alpha$ below and above $\alpha^*$ for  (a) DHL with $d=2$ and for values of $\alpha$ according to the legend colors. (b)  DHL with $d=3$ and for values of $\alpha$ according to the legend colors.}
 	\label{fig4}
 \end{figure}

The temperature dependence of the magnetization for values $0 \leq\alpha \leq \alpha_{\max}$ is illustrated in Figure \ref{fig4}.  For $0<\alpha \leq \alpha^* $ one continuous phase transition is achieved (for a fixed value of $\alpha$) the magnetization smoothly decreasing to zero at the critical temperature, while for $\alpha^* < \alpha \leq \alpha_{\max}$ two transitions are observed:  a sudden transition at low temperatures and a continuous one at higher temperatures. The latter describes the usual continuous transition from the \emph{disordered} paramagnetic to the ferromagnetic phase when the temperature is lowered from high temperatures. The former corresponds to a transition from the ferromagnetic phase characterized by a non-zero magnetization to the \emph{ordered} paramagnetic phase with zero-magnetization as the temperature is lowered. This  \emph{ordered} paramagnetic phase as discussed below is described by configurations with the majority of the spins in the state $S=0$, which differs from the usual high temperature \emph{disordered} paramagnetic phase where the configurations have, on the average, the fraction 1/3 of the sites randomly occupied by spins at states $S= 0, \pm 1$, respectively.

Figure \ref{fig4}(a) illustrates the temperature behavior of the magnetization for a DHL with $d=2$. The black plot  shows the special case $\alpha  = 0.0$ corresponding to the zero-fields spin-one Ising model where the expected continuous phase transition is achieved and is taken as reference. For $\alpha^*<\alpha \leq \alpha_{\max}$, however, two distinct behavior can be distinguished. For instance, the red graph was obtained for $\alpha_{\textrm d }\simeq 1.046$ corresponding to a low critical temperature ($T_{\textrm d}=0.060$) and exhibiting a full discontinuity in the magnetization (from zero to one), characteristic of first-order phase transitions (also observed for all values below  $\alpha_{\textrm d} \simeq 1.046$). This point of the transition line that corresponds to the maximum $\alpha$ value for which such strong discontinuity in magnetization is observed is marked with $\alpha_{\textrm d}$ becoming the candidate to be the lower bound of a \emph{tricritical} point. Moreover, for this value of $\alpha_{\textrm d}=1.046$ and for higher temperatures, a continuous transition from the ferromagnetic phase to the usual paramagnetic phase occurs, characterizing the re-entrance phenomenon. Furthermore for higher values of $\alpha$ ($\alpha_{\textrm d} < \alpha < \alpha_{\max}$) two plots are drawn obtained for $\alpha_{\textrm c} \simeq 1.151$ and $1.248$,  the blue and green ones, respectively. For both cases, each plot displays two phase transitions but with the magnetization varying continuously at their respective transition points. It is also important to note that the maximum value of magnetization in these latter plots decreases continuously to zero when $\alpha$ approaches $\alpha_{\max}$.
 
A similar scenario is observed for the magnetization in the DHL with dimension $d=3$, as shown in Figure \ref{fig4}(b). It is also shown the plots for the magnetization for the cases $\alpha_0 = 0.0$ and for three values with $ \alpha^*<\alpha  \leq \alpha_{\max}$. One difference is however noteworthy. The one with $\alpha = \alpha^*=1.500$ and the other with slightly higher $\alpha = 1,505$ behave quite differently. While in the first case the magnetization evolves continuously towards the value 1, in the latter a strong discontinuity to zero occurs, when the temperature is lowered, indicating the transition to zero magnetization phase. This high sensitivity of magnetization with respect to the $\alpha$ variation is due to the fact that the frontier, separating the ferromagnetic  and the zero magnetization phases, is vertical at low temperatures in the three-dimensional model, as it can be verified in Figure \ref{fig2}. 

The quadrupolar parameter $q$  \emph{per spin} is defined by
\begin{equation}
\label{eq38}
q=\frac{1}{N_s}\sum_{j=1}^{N_s}\langle \sigma^2_j \rangle,
\end{equation} 
which can be numerically calculated by iterating equations \eqref{eq27}-\eqref{eq31}  as a function of the temperature for fixed values of $\alpha$. This parameter combined with the magnetization order parameter $m$  can be used to obtain the mean fractions of sites $n_0, n_+,$ and $n_-$ in the states with $S = 0, + 1 $ and $ -1 $, respectively. It is reasonable to consider $n_0=1-q$, which measure the fraction of sites with spins at the state $S=0$, as the \emph{order parameter} of the \emph{ordered} paramagnetic phase associated with the conjugate field $\alpha=\Delta/J$ as usual \cite{jain80}. The magnetization per spin can be written as $m=(n_+-n_-)$ and the quadrupolar parameter $q=n_+ + n_-$, considering that $n_0+ n_+ + n_-=1$ we obtain  $n_\pm=(q\pm m)/2$. 

\begin{figure}[ht]
\begin{center}
\includegraphics[width=8cm]{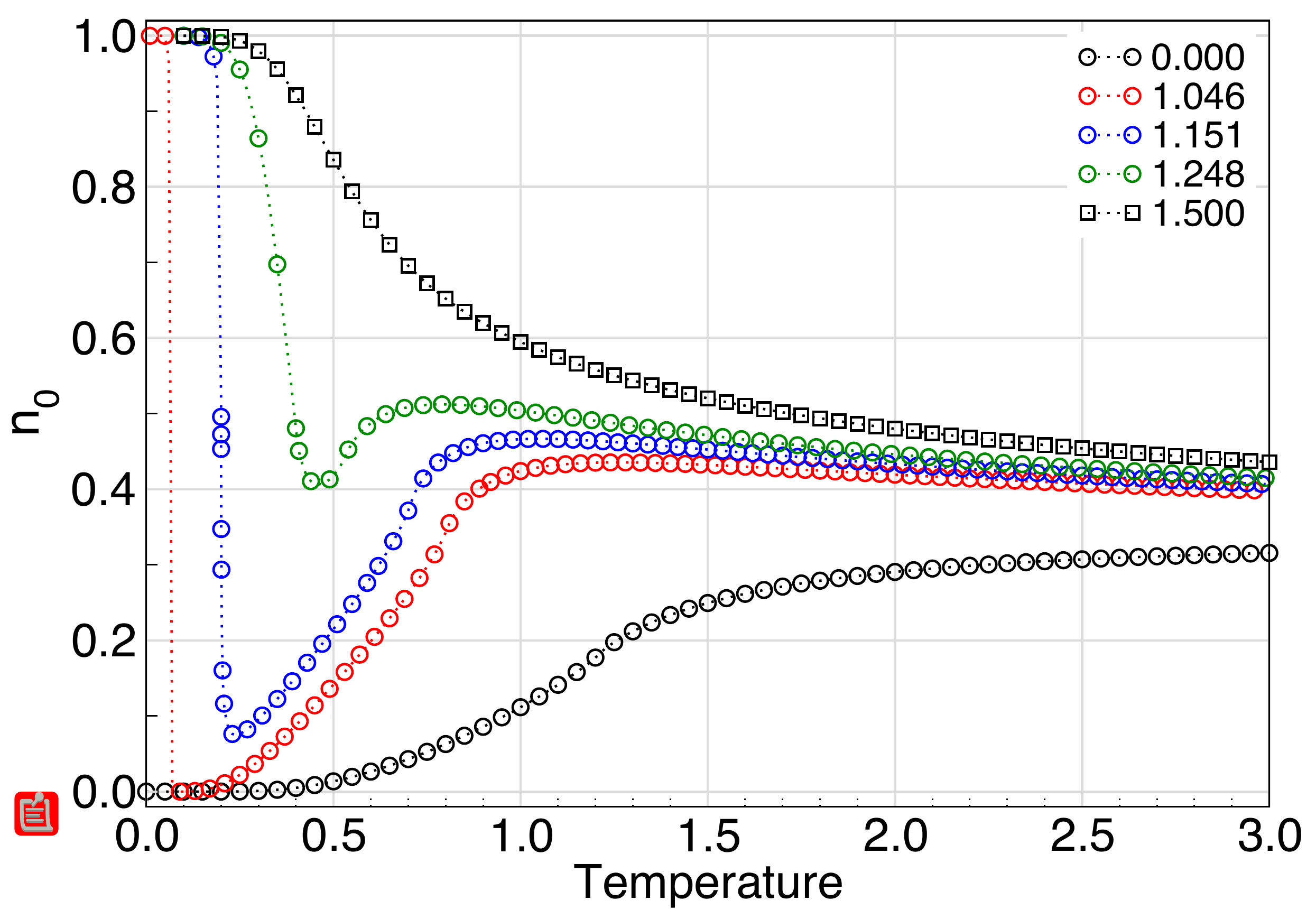}
\caption{Fraction $n_0$ of sites with spin configuration $S=0$  as a function of temperature for fixed values of the field parameter $\alpha$ as indicated in the legends.}
\label{fig5}
\end{center}
\end{figure}

The plots of $n_0$ against temperature for some fixed values of $\alpha \geq 0$ are  illustrated  in Figure \ref{fig5} for the $d=2$ DHL. The plot labeled with black circle symbols describes the dependence of $n_0$ with the temperature for $\alpha_0=0.0$ for the spin-one Ising model under zero crystal field, which is taken as a reference. We note that in this case, $n_0$ grows continuously from zero up to the 1/3 limit but with a growth rate reaching the maximum value at the transition point. The plots for $\alpha$ in the region with $\alpha^*<\alpha <\alpha_{\max}$ are labeled with colored circles whereas the one marked with black squares $\square$ was obtained for a typical value above $\alpha_{\max}$, all of them showing that $n_0 = 1$ at temperatures very close to zero. However, those with $\alpha^* <\alpha <\alpha_{\max}$ display an immediate and sharp decreasing of $n_0$ until reaching a minimum value followed by a smoother growth until reaching the 1/3 limit value at high temperatures. In all three cases, it is easy to verify through the numerical derivative that the minimum (maximum) points of the decrease (increase) rate signal the corresponding transition observed in the phase diagram. However, the plot obtained for $\alpha> \alpha_{\max}$ shows only a smooth decrease of $n_0$ from the maximum value 1 to the limit value of high temperatures. The behavior of $n_0$ can be better understood through its density plot.

\begin{figure}[ht]
\begin{center}
\includegraphics[width=8cm]{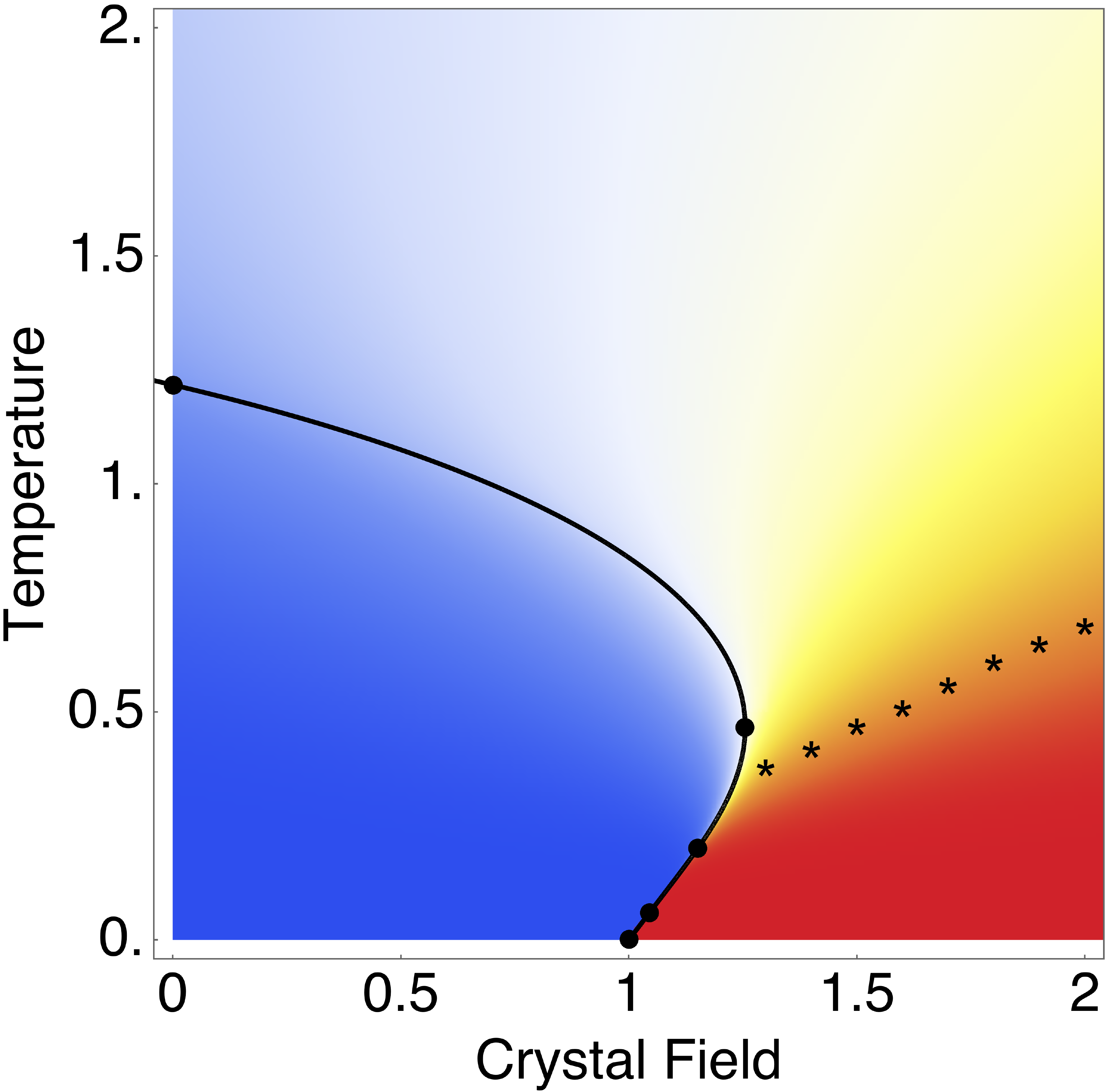}\quad 
\includegraphics[height=50mm]{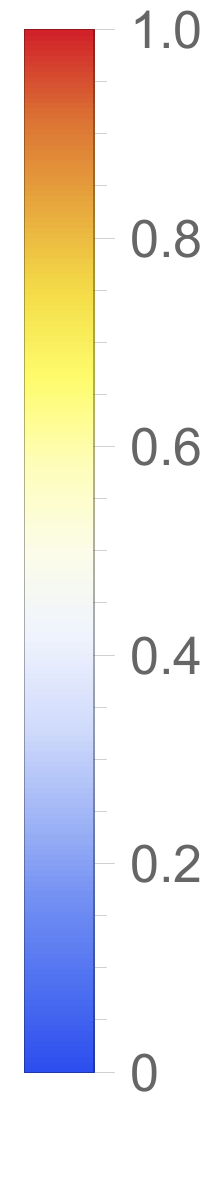}
\caption{Density plot of the fraction $n_0$ of spins at the state $S=0$ as a function of the temperature $(1/J)$ and the crystal field parameter $\alpha=\Delta/J$ for the DHL with dimension $d=2$. The continuous line locates the border separating the ferromagnetic phase to the paramagnetic phase. The stars ($\star$) mark the points where the derivative of $n_0$ with respect to the temperature reaches its minimum for a fixed value of $\alpha$ and  the dots ($\bullet$) mark the points $(T_0, \,\alpha_0)$, $(T_{\max},\,\alpha_{\max})$, ($T_{\textrm c},\,\alpha_\textrm c$), ($T_{\textrm d},\,\alpha_\textrm d$) and  $(T=0, \alpha^*$), from top to bottom as recorded in Table \ref{tab}. The colors set the density value according to the palette in the legend on the right.}
\label{fig6}
\end{center}
\end{figure}

\begin{figure}[ht]
\begin{center}
\includegraphics[width=8cm]{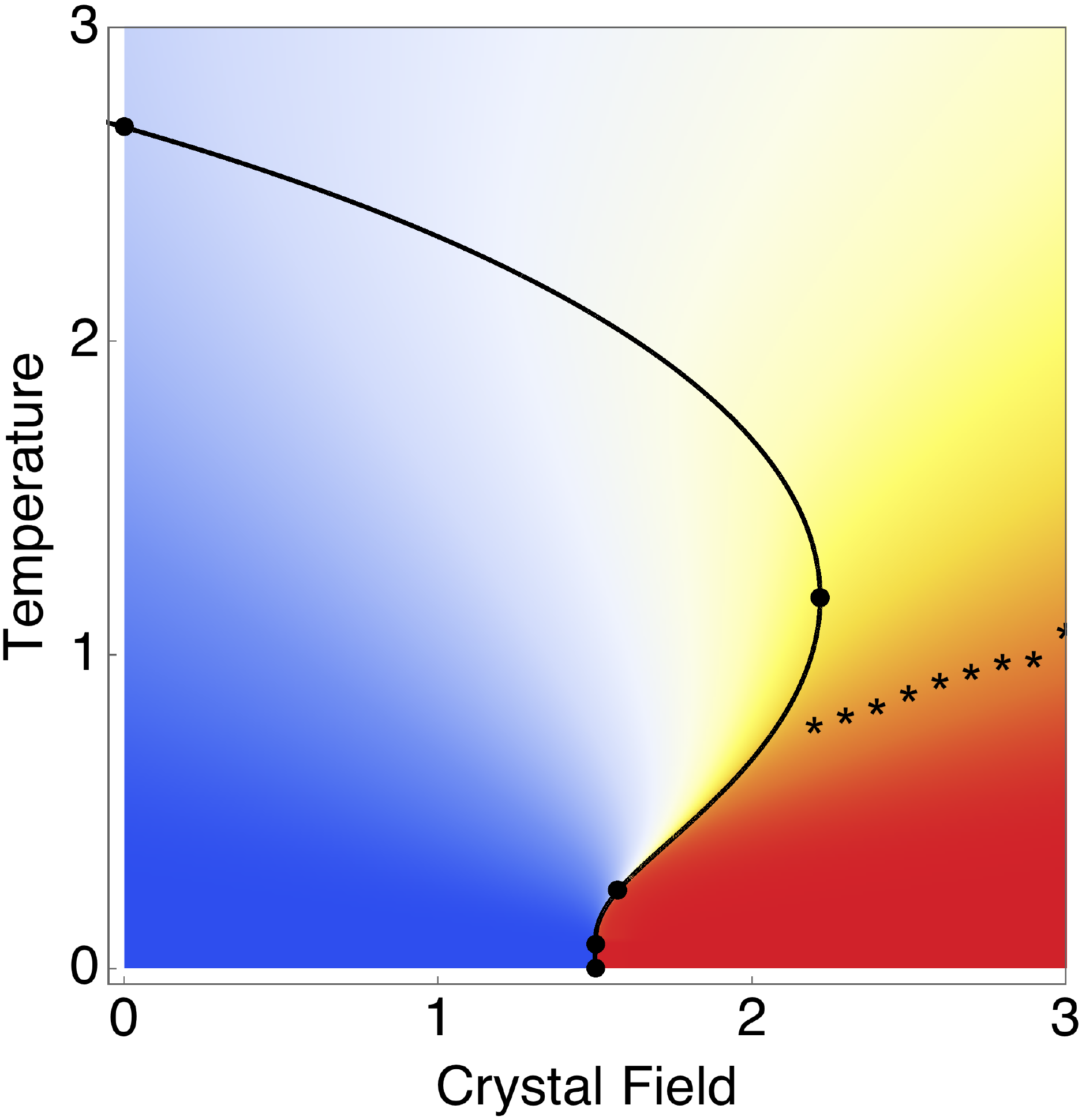} \qquad
\includegraphics[height=50mm]{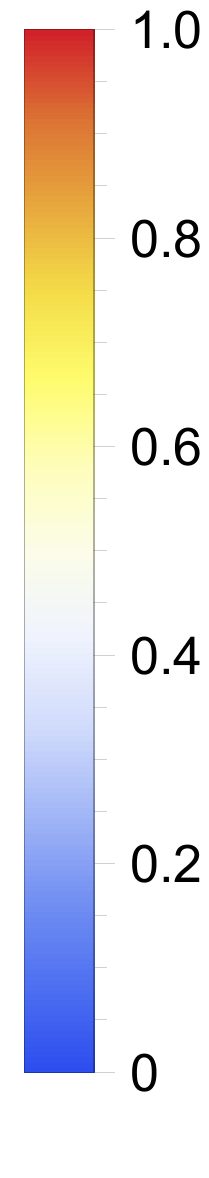}
\caption{Density plot of the fraction $n_0$ of spins at the state $S=0$ as a function of the temperature $(1/J)$ and the crystal field parameter $\alpha=\Delta/J$ for the DHL with dimension $d=3$. The continuous line locates the border separating the ferromagnetic phase to the paramagnetic phase. The stars ($\star$) mark the points where the derivative of $n_0$ with respect to the temperature reaches its minimum for a fixed value of $\alpha$ and  the dots ($\bullet$) mark the points $(T_0, \,\alpha_0)$, $(T_{\max},\,\alpha_{\max})$, ($T_{\textrm c},\,\alpha_\textrm c$), ($T_{\textrm d},\,\alpha_\textrm d$) and  $(T=0, \alpha^*$), from top to bottom as recorded in Table \ref{tab}. The colors set the density value according to the palette in the legend on the right.}
\label{fig7}
\end{center}
\end{figure}

The density plot of $n_0$ is illustrated in Figures \ref{fig6} and \ref{fig7} for DHL with dimension $d=2$ and $d=3$,  for the respective relevant regions of the phase diagram $T \times \alpha$. In both cases we also drew the borderline separating the ferromagnetic regions with non-zero magnetization to that with zero magnetization, the latter comprising the \emph{ordered} and the \emph{disordered} paramagnetic phase regions. It is important to note that for $\alpha$ values greater than $\alpha_{\max}$ the fraction $n_0$ of sites with spins in the $S = 0$ state decreases continuously from one to one-third (not appearing in the plot) when the temperature increases from zero to high temperatures as shown in Figure \ref{fig5}. The rate of this decreasing behavior, however, is not uniform.  Although the decay rate also reaches a minimum value there is no other sign of the existence of a phase transition in this case.  The locus of the points ($\alpha, T$) where the minimum occurs is marked with stars ($\star$) occurring for densities slightly below 0.9.   

\begin{figure}[ht]
\begin{center}
\includegraphics[width=10cm]{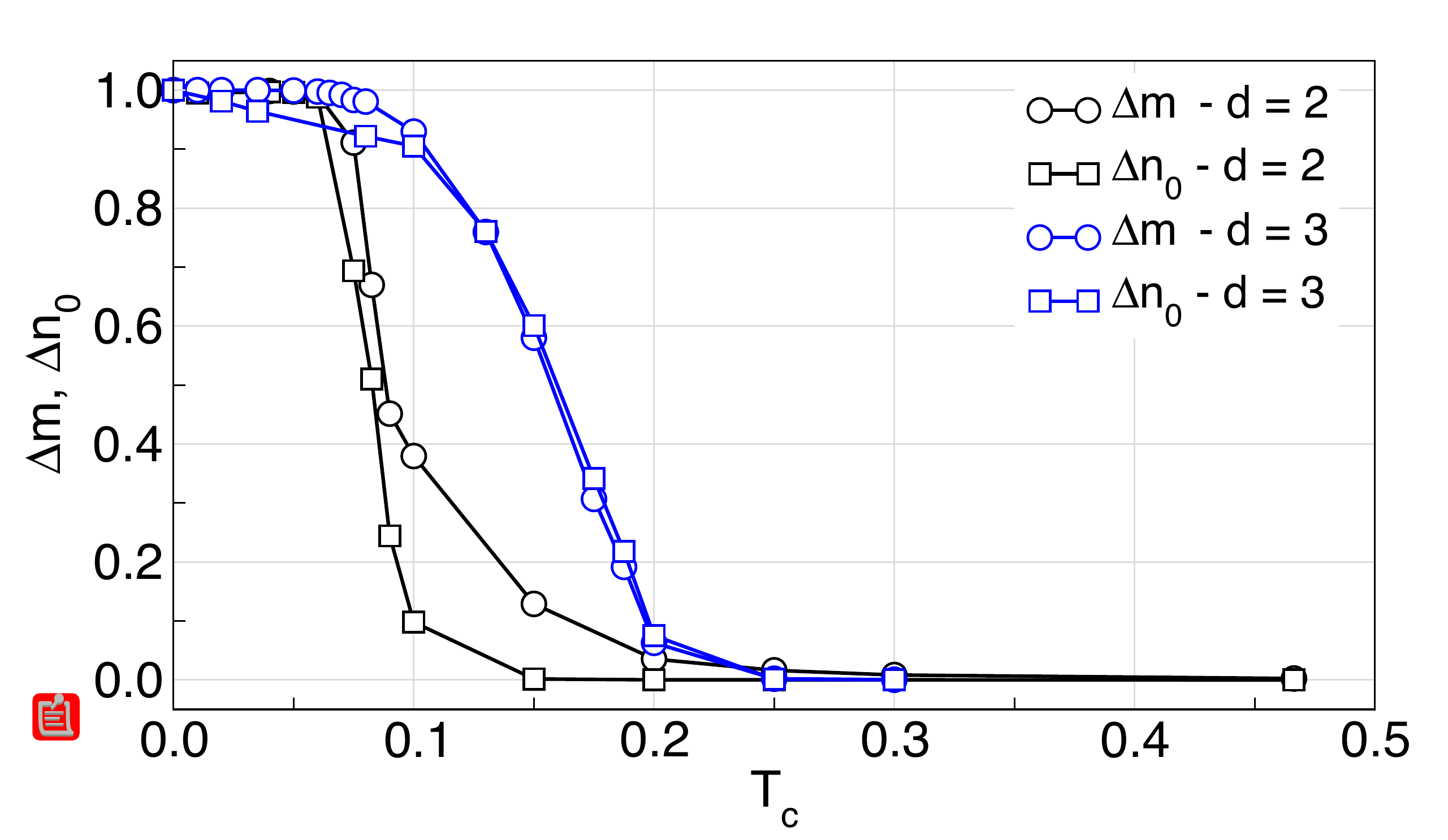} 
\caption{Discontinuities $\Delta m$ (open circles) of magnetization and  $\Delta n_0$ (open squares)  of the zero-magnetization order parameter along the frontier separating the ferromagnetic and zero-magnetization regions at low temperatures in the interval $\alpha^* \leq  \alpha <\alpha_{\max}$ for $d = 2$ (black symbols)  and $ d = 3 $ (blue symbols). }
\label{fig8}
\end{center}
\end{figure}

Next, we explore in detail the behavior of the model at low temperatures at the interface between the ferromagnetic and zero magnetization phases delimited by the interval $\alpha^* \leq \alpha \leq \alpha_{\max}$. For a fixed $\alpha$ value within this interval both the magnetization and the zero-magnetization order parameter $n_0$ exhibit an abrupt change when calculated at temperatures just below and slightly above the transition point, suggesting a first-order transition. The finite value of these discontinuities $\Delta m$ and $\Delta n_0$ vary from one to zero when calculated in the range of $\alpha_{\textrm d} \leq  \alpha <\alpha_{\textrm c}$ as illustrated in Figure \ref{fig8}. From the numerical data, we identified $\alpha_{\textrm d}\simeq 1.046$ and $\alpha_{\textrm c}\simeq 1.151$  for $d = 2$,  and $\alpha_{\textrm d}\simeq 1.500$ and $\alpha_{\textrm c}\simeq 1.555$ for  $d = 3$, as recorded in Table 1.

In the following section, the structure of the local magnetization very close to the transition in values of $\alpha_0 \leq \alpha \leq \alpha_{\textrm c}$ is investigated in detail with the support of the multifractal analysis.

\section{Local magnetization properties: multifractal analysis}
\label{section4}
\begin{figure}[ht]
\begin{center}
\includegraphics[width=100mm]{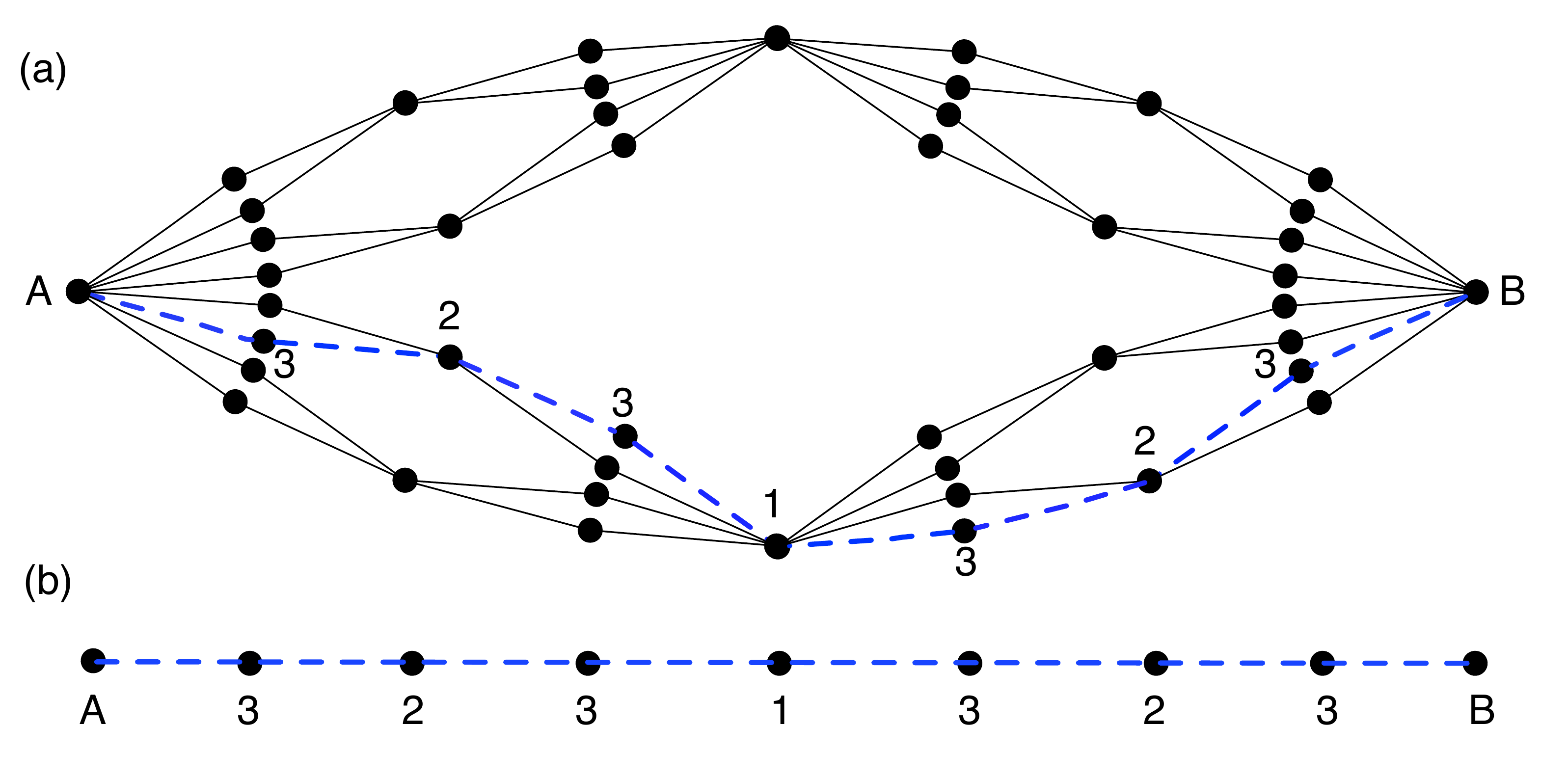}\\
\caption{Diamond Hierarchical Lattice with $n=3$ generations and fractal dimension $d=2$ (a) and the corresponding sites of the profile (b). The dashed blue line indicates a typical geodesic joining the root sites A and B. Numbers indicate the generation in which the sites were introduced.}
\label{fig13}
\end{center}
\end{figure}

The set of exact recursive equations (\ref{eq22}-\ref{eq26}) allows obtaining by iteration the magnetization at each site of a hierarchical lattice with $n$ generations. Values of this local magnetization can be better analyzed through its distribution along the sites of one of the shortest bond path (\emph{geodesic}) connecting the root sites of the DHL graph.  Figure \ref{fig13} shows a sketch of a finite DHL with $n = 3$ generations and fractal dimension $d = 2$ illustrating the sites belonging to a typical geodesic. The sequences of values obtained along any geodesic are \emph{identical}  due to the topological symmetry of the DHL graph. This representative sequence of values characterizes a \emph{profile} of the local magnetization in a given DHL. Each profile of a DHL comprises all kinds of sites having distinct coordination number and position with respect to the roots ones.  A \emph{profile} of a DHL with $n$ generations contains exactly $2^{n}+1$ representative sites. 

 \begin{figure}[ht]
\begin{center}
\includegraphics[width=90mm]{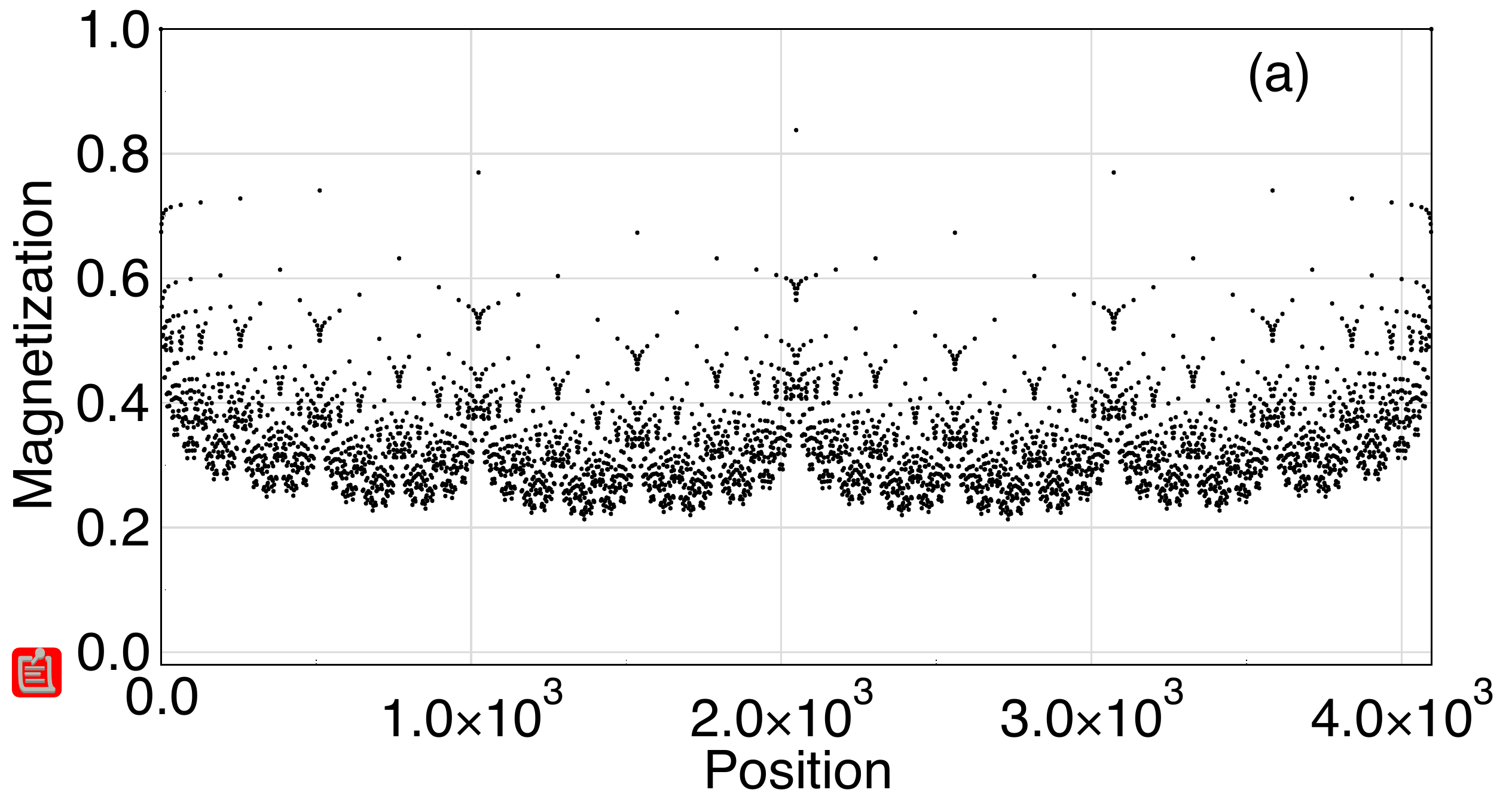}\\
\includegraphics[width=90mm]{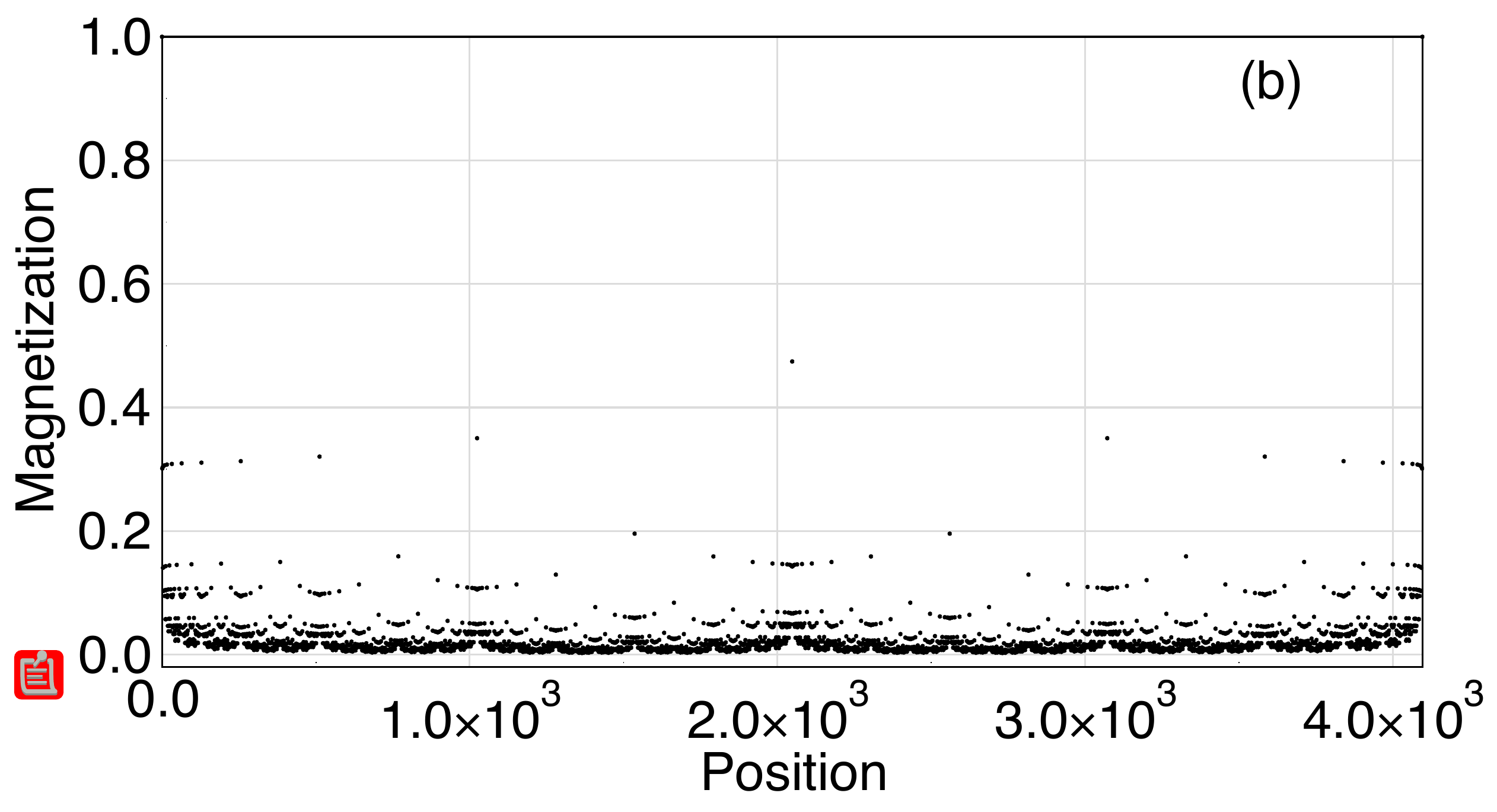}
\caption{Typical profiles of the local magnetization for the continuous transition of the zero crystal field BC model: (a) $d=2$ DHL  at $\alpha_0=0.000$ and $T_0 \simeq 1.217$; (b) $d=3$ DHL at $\alpha_0=0.000$ and $T_0 \simeq 2.682$. For the sake of clarity, both are shown with $n=12$ generations.}
\label{fig9}
\end{center}
\end{figure} 
In Figure \ref{fig9} we show the typical profiles of the magnetization in the continuous Ferro-Para transition for the case of the zero-fields spin-one Ising model at DHL with  $d=2$ and $d=3$, respectively.  Both profiles exhibit rich patterns with self-affine scale invariance reminiscent of the fractal structure of the DHL. For the DHL case with $d=2$ the average magnetization per site (of the profile) has a value $m = 0.264$, while for $d = 3$ this value is $m = 0.010$, while the corresponding values for the whole lattice are $m=0.244$ e $m=0.007$, respectively, the latter calculated in the limit of high generations. Note that the average magnetization per site for the entire lattice has a lower value than that for the corresponding profile. This happens because the magnetization of a site introduced in a higher $\ell$-generation has lower relative value and higher frequency ($p^{\ell}$) in the entire lattice. Hence contributing to the decrease of the average magnetization of the entire lattice with greater weight. However, the magnetization is expected to be zero in the continuous transition, which is not apparent in the case $d=2$ shown in Figure \ref{fig9}(a). However such expected limit $m \to 0$ must occur whenever $n\to \infty$, as analytically demonstrated in the reference \cite{morgado90} for the spin $\frac{1}{2}$ Ising model in the DHL.

\begin{figure}[ht]
\begin{center}
\includegraphics[height=50mm]{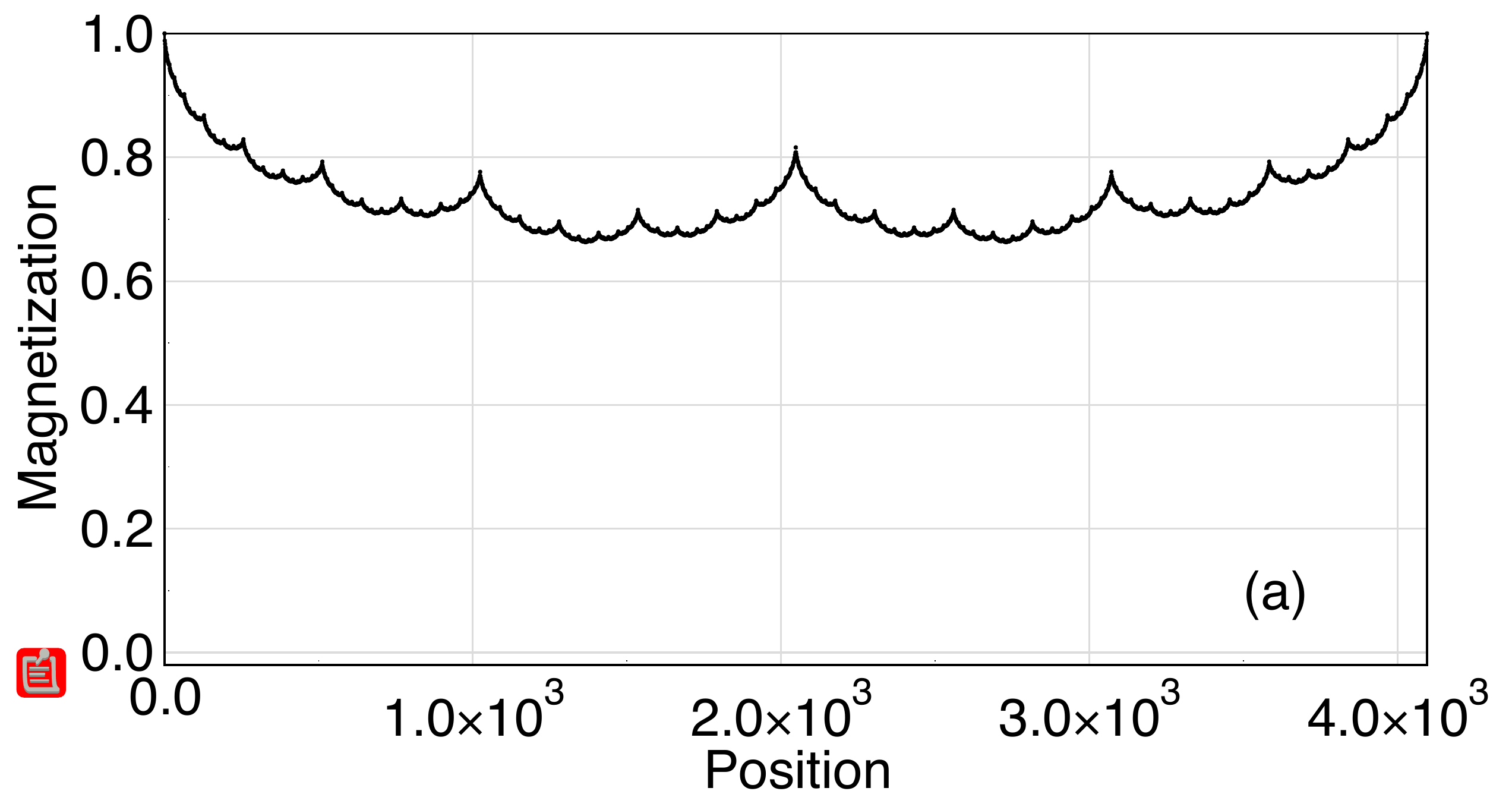}\\
\includegraphics[height=50mm]{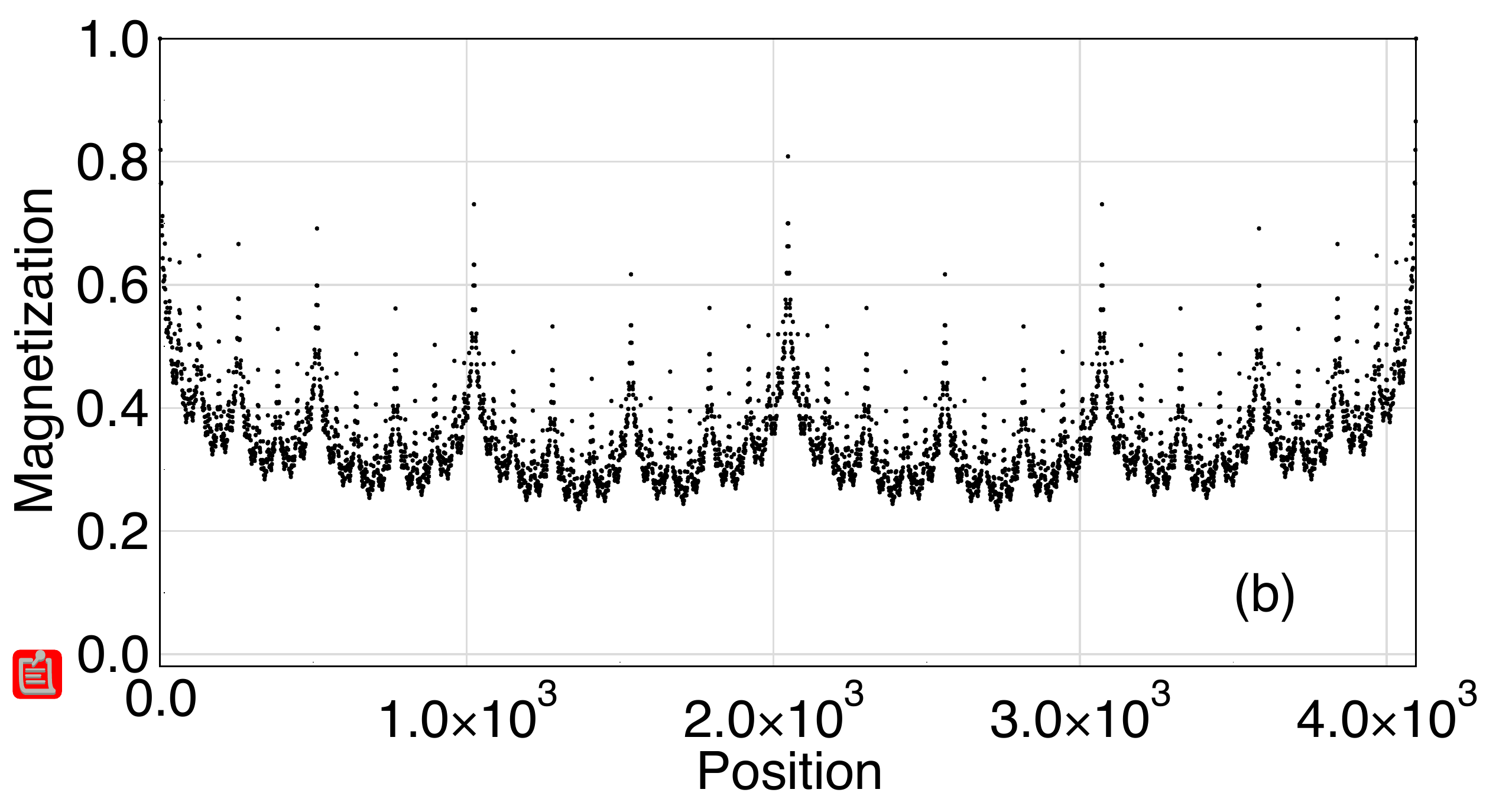}
\caption{Typical profiles of the local magnetization for the $d=2$ DHL  at (a): $\alpha=1.0764$ and $T\simeq 0.10 $ with average magnetization $m=0.386$; (b) $\alpha_{\textrm c}=1.150$ and $T_{\textrm c}\simeq 0.2 $ with average magnetization $m=0.186$. For the sake of clarity, plots are shown for a DHL with $n=12$ generations but the averages values were calculated for higher generations.}
\label{fig10}
\end{center}
\end{figure}

\begin{figure}[ht]
\begin{center}
\includegraphics[height=60mm]{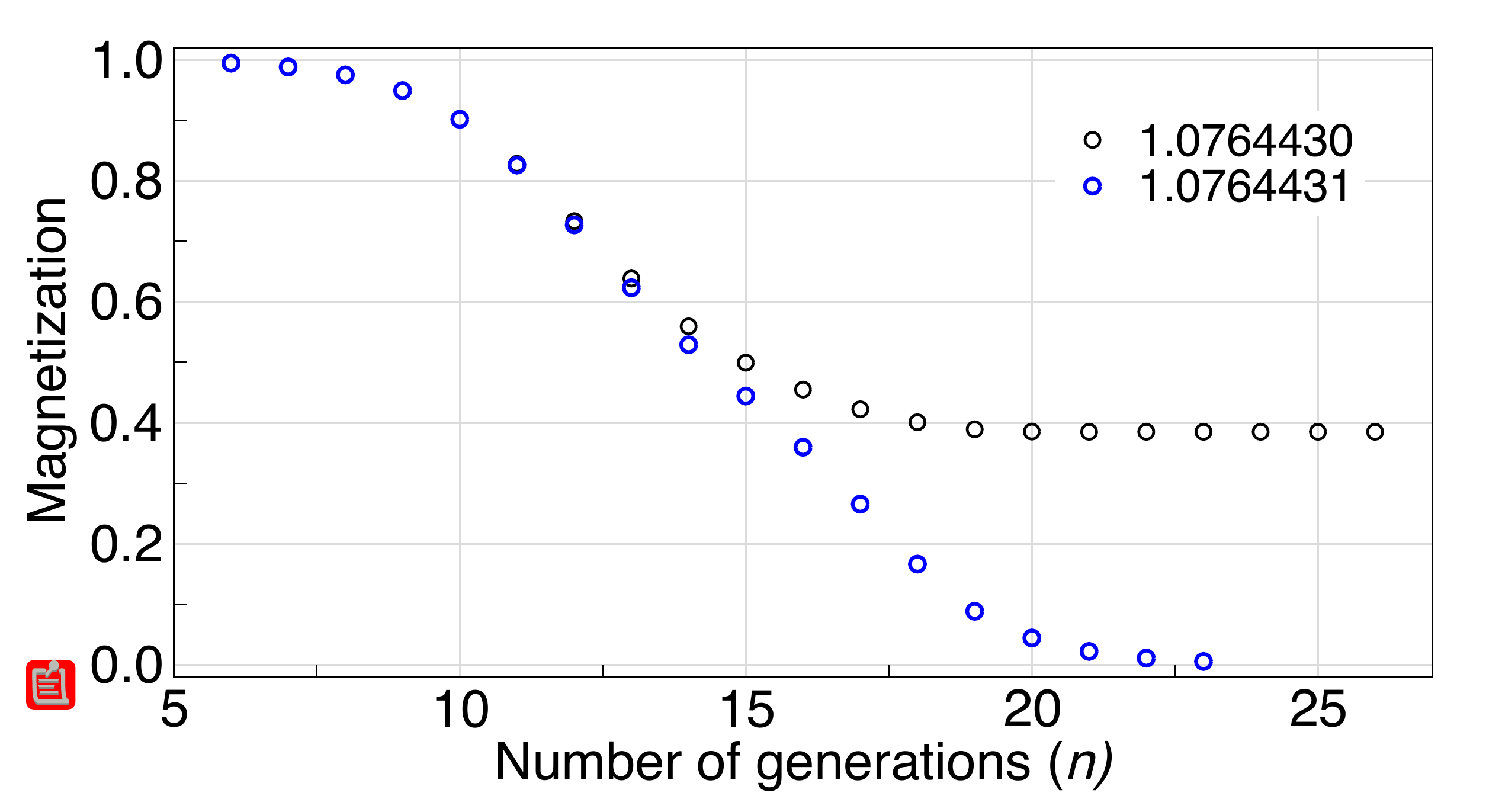}
\caption{Average magnetization versus the number of generations $n$ calculated for values of $\alpha$  slightly before and after the frontier between the  ferromagnetic and zero-magnetization regions at $T = 0.10$ for $d=2$ DHL (see legend).}
\label{fig11}
\end{center}
\end{figure}

In the low-temperature border line between the ferromagnetic and the ordered zero magnetization regions, the profiles exhibit quite different patterns from those mentioned above, as displayed in Figure \ref{fig10}. In these typical profiles, the structure of the magnetization becomes less irregular although they still present the self-affine scale invariance fractal characteristic. Moreover, they clearly exhibit finite and well defined average magnetization, which does not vanish as $n\to \infty$. This statement can be confirmed for instance by the plot of the mean magnetization versus $n$, calculated on the discontinuous transition line, as shown in Figure \ref{fig11} for the corresponding case of Figure \ref{fig10}(a).

\begin{figure}[ht]
\begin{center}
\includegraphics[height=70mm]{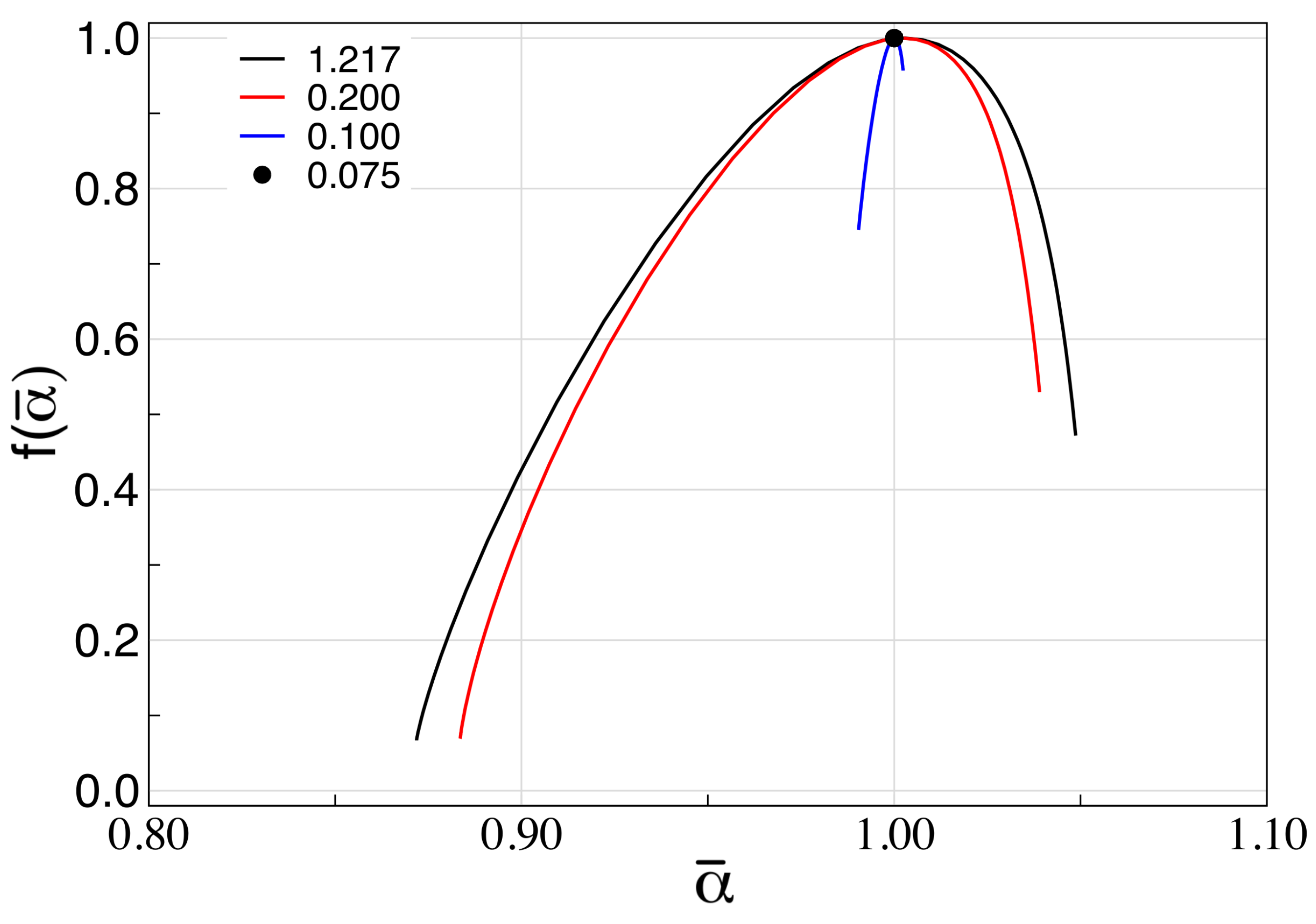}
\caption{Fractal spectrum for the patterns of the magnetization profiles at points on the transition line: black line at ($\alpha_0=0, T_0=2.217$); the red line for the continuous transition at ($\alpha_{\textrm c}=1.150, T_{\textrm c}= 0.200$); the blue line for the discontinuous transition at ($\alpha=1.0764, T=0.100$) and the black dot for $(\alpha_{\textrm d}=1.046, T_{\textrm d}=0.060)$ for the BC model with $d=2$. Colors indicate the corresponding critical temperatures according to the legend.}
\label{fig12}
\end{center}
\end{figure}

The main distinction between the patterns of the magnetization profiles shown in Figures \ref{fig9} and \ref{fig10} emerges through its respective \emph{multifractal spectra} $f(\bar\alpha)$, which here is calculated by the method due to Chhabra and Jensen\cite{chhabra89}.  First, we define the \emph{fractal measure} $\mu_i$ associated with local magnetization along the $N$ sites of the profile
\begin{equation}
\label{eq44}\mu_i=\ \frac{m_i}{\sum_i m_i}\,.
\end{equation}
To obtain the multifractal \emph{spectrum} or the function $f(\bar\alpha)$ we construct a $q$-parametrized family of normalized measures (probabilities) defined by
\begin{equation}
\label{eq45}
\bar\mu _i(q)=\frac{(\mu_i)^q}{\sum_i(\mu_i)^q}\mbox{,}
\end{equation}

\noindent
which is a generalization of the original measure $\mu _i$. $\tilde\mu _i(q)$ enhances the large (small) site probabilities for positive (negative) values of $q$. The  f$(\bar\alpha)$ spectrum is therefore obtained through the $q$-parametric function 
\begin{equation}
\label{eq46}f(\bar\alpha _q)=\lim _{N\to \infty }\left\{ \frac{-1}{N\ln 2}\sum^ N_i \bar \mu_i(q)\ln \bar\mu_i(q)\right\},
\end{equation}

\begin{equation}
\label{eq47}\bar\alpha_q=\lim _{N\to \infty }\left\{ \frac{-1}{N\ln 2}\sum_i^ N\bar\mu_i(q)\ln \mu_i\right\} \,.
\end{equation}
Figure \ref{fig12} displays the plots of the $f(\bar\alpha)$ spectra  for the profiles shown in Figures \ref{fig9}(a) and \ref{fig10}(a,b) and the additional dot $(1,1)$ which represent the \emph{trivial} spectrum for the \emph{fractal} profile of the magnetization at $T_\textrm{d}=0.060$ ($d=2$ DHL). We notice from this figure that at the continuous transitions the spectrum is much wider than the one where a discontinuous transition occurs, suggesting that in the latter case the profile approaches a simple fractal pattern while for the former it presents a well-defined multifractal pattern. Therefore a multifractal spectral pattern (or its absence) can be used also as a criterion for deciding the nature of the transition on the frontier that separates the ferromagnetic and ordered paramagnetic phases. 

\begin{figure}[ht]
\begin{center}
\includegraphics[height=70mm]{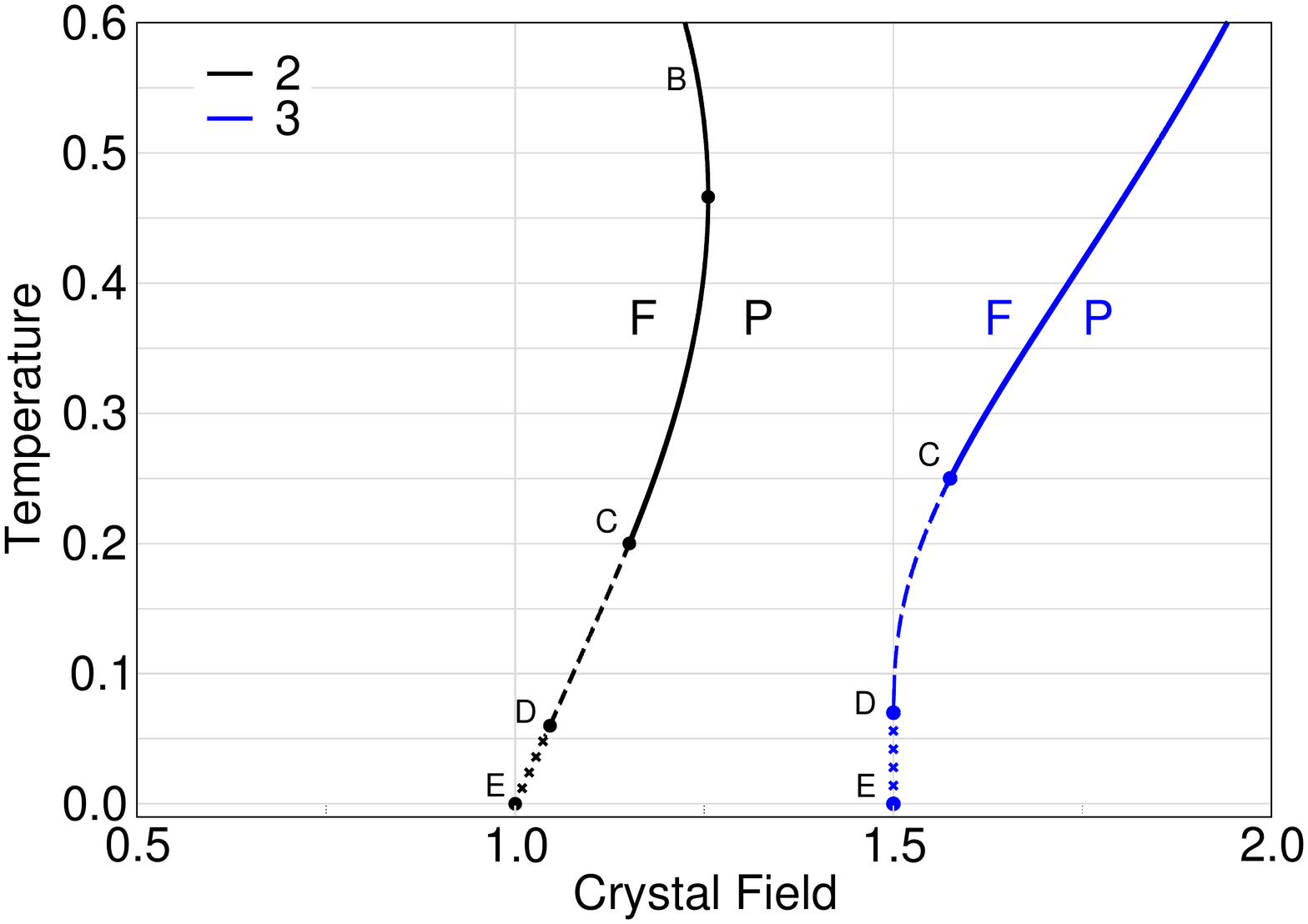}
\caption{Temperature ($1/J$) versus the crystal field parameter ($\alpha= \Delta/J$) at low temperatures for BC model defined in the DHL with dimensions $d=2$ (black lines) and $d=3$ (blue lines). F and P denote the corresponding ferromagnetic and paramagnetic regions, respectively. B, C, D and E label the points reported in Table \ref{tab}. The solid line indicates a continuous phase transition.  The Xs mark a line of first-order phase transition with a full discontinuity in the order parameter and a fractal profile of the magnetization. The dashed line delimits a discontinuous phase transition with a finite order parameter discontinuity and multifractal profile of the magnetization.}
\label{fig14}
\end{center}
\end{figure}

The first-order phase transition frontier between the ferromagnetic and the ordered paramagnetic phases of the BC model can be drawn based on such criterion and on the observation of full discontinuities of the order parameters. We found the upper bounds points D for this frontier: $(T_\textrm{d} = 0.060, \alpha_\textrm{d} = 1.046)$ for the case $d =2$ and ($T_\textrm{d} =0.070, \alpha_\textrm{d} =1.500$) for the case $d = 3$, respectively. Figure \ref{fig14} displays a zoom of the phase diagram, temperature ($1/J$) versus the crystal field parameter ($\alpha= \Delta/J$), at low temperatures for BC model defined in the DHL with dimensions $d=2$ (black lines) and $d=3$ (blue lines). In these plots,  B, C, D and E label the points reported in Table \ref{tab}. The solid lines indicate the continuous phase transitions while the dash-dotted lines (D--E) mark the first-order phase transition characterized by a full discontinuity in the order parameter and a fractal profile for the local magnetization. The dashed lines (C--D) delimit, however, a discontinuous phase transition which is characterized by (a) a finite order parameter discontinuity that vanishes monotonically along the line as the temperature grow and (b) a multifractal profile pattern for the local magnetization that enhances the width of its spectrum from zero (fractal) up to a finite one observed in second-order phase transitions. We expect that the locus of a possible \emph{tricritical} point of the BC model defined in DHLs shall be located along this line (C--D).

\section{Discussion and Conclusions}\label{section5}
In this work we studied the local properties of the order parameters of the ferromagnetic BC model in dimensions two and three, at low temperatures and zero external field, associated with the condensed phases with non-zero magnetization (ferromagnetic) and zero magnetization (ordered paramagnetic), the latter with predominance of sites in the $S = 0$ state. We focused our study on the behavior of the magnetization $m$ and the density $n_0$ of sites with state $S = 0$, as functions of the temperature and the anisotropic crystal field, calculated by an exact procedure for the model defined on diamond hierarchical lattices. 

The results were inspected according to the values of the parameter $\alpha=\Delta/J$ which measure the strength of anisotropic crystal field coupling with respect to the exchange interaction constant between spins according to
$\alpha < \alpha^*$, $\alpha^* \leq \alpha < \alpha_{\max}$ and $\alpha \geq \alpha_{\max}$. 

In the interval $\alpha < \alpha^*$ the BC models with dimension $d\geq 2$ presents solely a continuous phase transition from the low-temperature ferromagnetic phase to the disordered paramagnetic phase exhibiting a broad multifractal spectrum for the magnetization profile at the transition. This interval includes the special cases $\alpha = 0$, which corresponds to the spin 1 Ising model (in the absence of the fields), and the $\alpha \to -\infty$ limit corresponding to the spin 1/2 Ising model. 

The second interval $\alpha^* \leq \alpha < \alpha_{\max}$ presents a reentrant phenomenon and a very rich physical behavior. When the temperature is lowered from higher values, the system undergoes a continuous phase transition from the \emph{disordered} paramagnetic phase, characterized by disordered configurations with on average one-third of the sites in each of the possible states $S = 0$ and $\pm 1$, to the ferromagnetic phase. When the temperature is lowered further we observed an increase of the magnetization followed by an unexpected phase transition from the \emph{ordered} ferromagnetic phase to the \emph{ordered} paramagnetic phase, the latter characterized by zero magnetization and the majority of spins in the state $S = 0$, i.e. $n_0\simeq 1$. In this latter transition, we observed the occurrence of high discontinuities in the magnetization and in the density $n_0$ for certain $\alpha$ values ($\alpha^* <\alpha < \alpha_\textrm{d}$), characteristics of a first-order phase transition. For $\alpha$ values just above $\alpha_\textrm{d}$ and temperatures on the transition line, finite discontinuities in $m$ and $n_0$ are still observed however vanishing as $\alpha$ grows up to $\alpha_\textrm{c}$, signalizing a possible change in the nature of the transition. 

To analyze the behavior of $m$ in this range of ($\alpha$, $T$) values we considered the profile of the local magnetization along a \emph{geodesic} between the root sites of the lattice graph. These profiles have a multifractal structure with a finite broad multifractal spectrum for continuous transitions as showed for the special case when $\alpha=0$ (spin-one Ising model) and in several other spin models defined in hierarchical lattices\cite{morgado91,coutinho92, silva96}. Typical profiles and respective multifractal spectra were achieved for $\alpha > \alpha_\textrm{d}$ values on the transition line at points where finite discontinuities occur in $m$ and $n_0$. We observed that when the temperature (and also $\alpha$) is lowered along the transition line from $(\alpha_\textrm{c}, T_\textrm{c}$) toward the point $(\alpha_\textrm{d}, T_\textrm{d}$) the profiles become less and less irregular and the corresponding multifractal spectra become narrower converging upon to a single dot that is characteristic of simple fractal. This behavior that is observed in both models for $d = 2$ and $d = 3$ could be taken as an indicator of the location of the tricritical point. However, the corroboration and unambiguous characterization of a tricritical point on line D--C require the calculation of its critical exponents and the analysis of the thermodynamic behavior of the model demanding a careful numerical effort, which is in progress to be presented shortly in a future work.

Finally,  the region of values of $\alpha> \alpha_{\max}$ does not show any phase transition at finite temperatures. However, the model has an \emph{ordered} and \emph{non-degenerate} ground-state, which is characterized by the configuration of all spins in the state $S = 0$ ($n_0=1$). For a fixed value of $\alpha$, the density decreases continuously when the temperature increases however nonlinearly. It exhibits a rapid decrease and then softens until it reaches the one-third limit value. 

\section*{Acknowledgments}
This work received financial support from CNPq and CAPES (Brazilian agencies for research support). M.J.G. Rocha-Neto thanks CNPq for the scholarship received under the grant 141185/2013.9\, .

\end{document}